\documentclass[aps,twocolumn,showpacs,superscriptaddress,groupedaddress,nofootinbib]{revtex4}  
\pdfoutput=1 

\usepackage{graphicx}  
\usepackage{dcolumn}   
\usepackage{bm}        
\usepackage{amssymb}   

\usepackage{xcolor}
\definecolor{myOrange}{rgb}{1,0.5,0}
\colorlet{myPurple}{blue!40!red}

\newcommand{\be}{\begin{equation}}
\newcommand{\ee}{\end{equation}}
\newcommand{\beq}{\begin{eqnarray}}
\newcommand{\eeq}{\end{eqnarray}}

\def\nue{\mathrel{{\nu_e}}}

\def\barnue{\mathrel{{\bar \nu}_e}}

\newcommand{\n}{neutrino}
\newcommand{\ns}{neutrinos}
\newcommand{\gw}{GW}
\newcommand{\lgvg}{LIGO-Virgo}
\newcommand{\nsns}{BNS}
\newcommand{\nsbh}{NSBH}
\newcommand{\mm}{multi-messenger}
\newcommand{\ck}{Cherenkov}
\newcommand{\bh}{black hole}

\newcommand{\mg}{merger}
\newcommand{\mgs}{mergers}

\newcommand{\bi}{\begin{itemize}}
	\newcommand{\ei}{\end{itemize}}

\newcommand{\eq}{Eq.}

\newcommand{\fig}{Fig.}

\newcommand{\Sec}{Sec.}

\newcommand{\equ}[1]{\eq~(\ref{equ:#1})}
\newcommand{\figu}[1]{\fig~\ref{fig:#1}}

\begin{document}



\title{Observing cosmological binary mergers with next generation neutrino and gravitational wave detectors}
\author{Zidu Lin\footnote[1]{\makebox[1.cm]{Email:}
		zlin28@asu.edu} }

\affiliation{Department of Physics, Arizona State University, \\ 450 E. Tyler Mall, Tempe, AZ 85287-1504 USA}
\author{Cecilia Lunardini\footnote[2]{\makebox[1.cm]{Email:}
		Cecilia.Lunardini@asu.edu}}

\affiliation{Department of Physics, Arizona State University, \\ 450 E. Tyler Mall, Tempe, AZ 85287-1504 USA}

\date{\today}

\begin{abstract}
We discuss the potential of detecting thermal \ns\ from matter-rich binary mergers, via a decades-long  multi-messenger campaign involving a  Mt-scale water \ck\ \n\ detector and one or more next generation gravitational wave detectors, capable of observing \mgs\ up to redshift $z\sim 2$.  The search of \ns\ in time-coincidence with gravitational wave detections will allow to identify single \ns\ from individual mergers above the background, and to study their distributions in energy, redshift and type (double neutron-star or neutron-star-black hole merger) of the candidate sources. We find that, for merger rates consistent with current \lgvg\ constraints, and for a $100~{\rm Mt\cdot yr}$ exposure, between ${\mathcal O(10^{-1})}$ and ${\mathcal O(10)}$ \n\ events are expected. For  extreme cases of mergers with more than $10^{52}$ ergs emitted in $\barnue$, the number of events can be as large as $\sim 100$, with sensitivity to mergers up to redshift $z\sim 0.5$ or so.  Such scenarios can already be tested with a $10~{\rm Mt\cdot yr}$ exposure, resulting in constraints on the post-merger evolution of the systems being considered.  
\end{abstract}

\maketitle

\section{\label{sec:intro}Introduction}

 Gravitational wave (\gw) astronomy is rapidly developing. Since the first discovery of \gw\ from a binary black hole merger in 2016 \cite{Abbott:2016blz}, several binary mergers of black holes and neutron stars have been observed by \lgvg\ (see, e.g. \cite{LIGOScientific:2018mvr}).  One event in particular, a binary neutron star merger, was heralded as the first \mm\ observation of a \mg, with detections in \gw\ and at electromagnetic frequencies \cite{TheLIGOScientific:2017qsa,GBM:2017lvd,Monitor:2017mdv,Soares-Santos:2017lru,Hallinan:2017woc}.  It delivered a wealth of scientific information, largely confirming theoretical predictions of the merger dynamics and \gw\ emission, and establishing neutron star mergers as sources of short gamma ray bursts \cite{Monitor:2017mdv} and sites of r-process nucleosynthesis \cite{Drout:2017ijr,Chornock:2017sdf,Pian:2017gtc}.   Several open questions remain, especially about the nature of the compact object which is born in the merger and its short- and long-term evolution.  

At this time, \ns\ are still missing from the \mm\ picture of \mgs\ \cite{ANTARES:2017bia,Abe:2018mic,Albert:2018jnn,Avrorin:2018fzl}.  Neutrino production is expected in compact-object mergers, i.e., binary neutron star (\nsns) and neutron-star-black hole (\nsbh) mergers. One site of \n\ production is the remnant of mergers and its accretion disk, where the dense and hot conditions allow efficient cooling via \ns\ similarly to a core collapse supernova.  Therefore, a burst of $\sim$10 MeV, thermal \ns\ is a generic prediction (see, e.g. \cite{Janka:1995cq,Ruffert:1996by,Rosswog:2003rv}). Furthermore, another possible channel of \n\ emission is due to the pion and muon decays generated through the photomeson production. The latter takes place when accelerated cosmic ray particles propagate through background photons and produce high energy (multi-TeV) neutrinos \cite{Gao:2013mcx,Fang:2017tla,Kimura:2017kan,Murase:2017snw,Biehl:2017qen}.  This scenario predicts a high energy \n\ flare accompanied by a gamma ray burst. Strong \n\ emission is possible if  the remnant is a (quasi-)stable magnetar, whereas little or no high energy \n\ production is expected if a black hole is directly produced (e.g., \cite{Gao:2013mcx,Fang:2017tla,Kimura:2017kan}).  

The theme of this paper is the thermal neutrinos from \nsns\ and \nsbh\ \mgs\ and their detectability.  A detection of these \ns\ would be very important to advance our understanding of \mgs. Indeed, while \gw\ data are mostly sensitive to the physics before and during the merger, \ns\ can give unique information on the post-merger phase. Models show that the duration of the \n\ burst, its luminosity and its energy spectrum depend strongly on 
the type of remnant produced (stable neutron star, transient neutron star or black hole), on the mass and accretion rate of the accretion disk surrounding the remnant, etc. (see e.g., \cite{Janka:1999qu,Sekiguchi:2011zd,Just:2014fka,Fujibayashi:2017xsz,Lippuner:2017bfm} for examples, more details in Sec. \ref{sub:models}). The features of the \n\ burst in turn have an important impact, through their effect on the electron fraction, on the  rates of formation of heavy elements via the r-process, see, e.g., \cite{Wanajo:2014wha,Foucart:2015gaa,Roberts:2016igt}.
Moreover, studying the \n\ emission could be important to understand the process of neutrino-antineutrino   annihilation into $e^+\ e^-$ pairs. This mechanism transfers part of the gravitational energy of the accretion disk  to electromagnetic radiation, thus possibly powering a short gamma ray burst \cite{Dessart:2008zd,Janka:1999qu,Fujibayashi:2017xsz}. 
Finally, the flavor energy  spectra of the merger \n\ burst at Earth will carry an imprint of the complex pattern of \n\ flavor oscillations inside the accretion disk, with the potential to probe its density and temperature profiles (e.g., \cite{Malkus:2015mda,Zhu:2016mwa,Frensel:2016fge,Wu:2017qpc,Tian:2017xbr}).

The thermal \ns\ from  \mgs\ are very challenging to detect. Their diffuse flux  is overwhelmed by the diffuse supernova \n\ background -- which has a similar spectrum -- and by experimental backgrounds; only in extreme cases, its observation might be possible \cite{Caballero:2009ww,Schilbach:2018bsg}.  Recently, a long-term strategy has been proposed by Kyutoku and Kashiyama \cite{Kyutoku:2017wnb}, which is based on exploiting the time-coincidence with \gw\ detections to strongly reduce backgrounds, and identify individual \n\ events from mergers.   Assuming the projected performance of the future Advanced LIGO \cite{Abadie:2010cf}, with its sensitivity up to $\sim 200$ Mpc distance (i.e., redshift $z\simeq 0.05$), it was found that a Mt-scale water \ck\ detector like the planned HyperKamiokande \cite{Abe:2011ts,Abe:2018uyc} might obtain a high significance detection of ${\mathcal O}(1)$ \ns\ from mergers, over about a century of operation.  The operating time needed would be significantly reduced if a multi-Mt detector like DeepTITAND \cite{Kistler:2008us} and MICA \cite{Boser:2013oaa} is constructed in the next few decades.

In parallel to the development of \n\ observatories, next generation \gw\ detectors, like the proposed Voyager \cite{T1600119}, Einstein Telescope \cite{Punturo:2010zz}, and Cosmic Explorer \cite{Evans:2016mbw}, will become a reality. They will allow a high-efficiency sensitivity to binary mergers as far as $z\sim 2$  \cite{Mills:2017urp}, thus opening up a much larger cosmic volume for searches of \ns\ in coincidence with \gw\ detections.  Inspired by this prospect, here we elaborate on the time-coincidence method for a future scenario where both a Mt-scale \n\ detector and at least one \gw\ detector with cosmological ($z\gtrsim 1$) sensitivity  are operating in synergy for several decades.   One main element of novelty of this work is that we examine the detection of \ns\ produced at cosmological distances, studying the contributions of mergers at different redshifts to the total signal.
We also discuss the significance and physics potential of a very low statistics signal, at the level of a single neutrino per decade or so. Another new element is that results are shown for different assumptions on the merger rate, and for several different models of \n\ emission from mergers. 

The paper is structured as follows: in Sec. \ref{sec:flux}, scenarios of \n\ production in \nsns\ and \nsbh\ mergers are illustrated, and the calculation of the \n\ flux at Earth is presented. In Sec. \ref{sec:rates}, the method of detection is described, and results are shown for the number of events of signal and background, and for the probability of detection as a function of the detector exposure.  In the discussion section, Sec. \ref{sec:disc}, the results are examined in their broader implications, and conclusions are drawn. 

\section{\label{sec:flux}Neutrinos from binary mergers: fluxes and spectra}

\subsection{\label{sub:models}Neutrino emission in mergers}

We consider mergers of binary systems made of two neutron stars or a neutron star and a stellar-mass black hole. 
We assume that the black holes are of stellar origin, i.e., the product of core collapse supernovae. 

As a result of a \mg, a massive remnant is formed, surrounded by an accretion disk. For \nsbh\ \mgs, the outcome is a black hole-torus system. 
Matter in the torus is accreted on the black-hole, so that
part of its gravitational binding energy is converted into internal energy. For a mass $M_{accr} \sim 0.1 M_\odot$ of material being accreted, the internal energy budget is $E_{tot} \gtrsim 10^{53}$ erg,  of which about half is thermal energy and is released via neutrino emission \cite{Dessart:2008zd,Perego:2014fma}. 
For \nsns\ \mgs, the post merger phase could be different depending on the binary parameters and the properties of the nuclear equation of state. It could lead to the direct formation of a \bh,
in which case \n\ production would proceed similarly to the \nsbh\ case. 
Other possibilities are either the formation of a stable NS in the central remnant or the birth of a transient hypermassive neutron star (HMNS). The latter might eventually collapse into a black hole. 
In the presence of a newly formed neutron star, the neutrino luminosity receives contributions  from both the accretion process and the cooling of the neutron star itself. The neutrino luminosity from a HMNS or a stable neutron star is powered by an internal energy reservoir, which is approximately the difference between the internal energy of a hot and of a cold HMNS/NS. For a HMNS of $\sim  2.5 M_\odot$ and $\sim 15$ MeV , the internal energy  is  $E_{th} \sim 3\times\ 10^{52}\ $ ergs \cite{Perego:2014fma}.

For both \nsns\ and \nsbh\ mergers, typically the \n\ emission is dominated by $\nue$ and $\barnue$,  with a certain preponderance of $\barnue$  due to the matter of the  disk -- and of the compact object, if a neutron star is formed -- being neutron-rich.  
In scenarios where a HMNS collapses into a black hole, the $\nue$ and $\barnue$ 
luminosities only decrease gradually, around the time of collapse, because these species continue to be produced in the accretion disk via charged current processes. On the other hand,
the muon and tau neutrino luminosity decreases sharply after the black hole formation because their inner, hotter surface of last scattering is swallowed by the black hole. 
The duration of a  \mg\ \n\  burst may  be of the order of a fraction of a second, up to several seconds, depending on the time-scale of the evolution of the central object \cite{Sekiguchi:2011zd,Sekiguchi:2015dma,Janka:1999qu,Just:2014fka,Fujibayashi:2017xsz}. 

A detailed description of the physics of \mgs, and their \n\ emission requires large scale numerical modeling. Here we adopt a simplified description of a burst of  $\barnue$ with thermal (Fermi-Dirac) spectrum with duration of up to $\sim 1$s.  The effect of flavor oscillations on the $\barnue$ spectrum and luminosity at Earth will be accounted for by a energy-independent suppression factor (see Sec. \ref{sec:rates}).
The total energy emitted, as well as the burst duration and average energies are drawn from numerical simulations; as summarized in Table   \ref{tab:models}. 
These simulations suffer from various approximations that are made necessary by the limited computational resources. Still, they are state-of-the-art, and give reasonable estimations on ejected matter as well as r-process nucleosynthesis properties consistent with previous works \cite{Sekiguchi:2016bjd,Martin:2015hxa,Kyutoku:2013wxa,Kyutoku:2017wnb}. Therefore, the  model results in Table  \ref{tab:models} are credible, and  can be used to
define a space of the parameters that can be considered plausible for \ns\ from mergers. Here, we will focus mainly on the two-dimensional space of the  total energy  in $ \barnue$ and the average energy of the $\barnue$ spectrum (see Fig. \ref{fig:Contour}).

A brief description of each model follows below.


	\begin{table*}
		\centering
		\caption{ Parameters and neutrino emission properties for the merger models considered in this study. The columns from left to right contain: (1) time duration, $T_{99},$ within which $99\%$ of the total energy in $\barnue$ is emitted; (2)  total energy emitted in $\barnue$; (3) average energy of the $\barnue$ (time-integrated) flux; (4) type of merger (\nsns\ or \nsbh); (5) type of central remnant (CR) which could be a hypermassive neutron star (HMNS) or a black-hole (BH); (6) the central remnant mass (CRM) in units of solar mass, $M_{\odot}$; (7) the disk mass (DM; note that for model M of Sekiguchi et al., this quantity is not available); (8) the model name and (9) corresponding reference (first author only; see text for full reference). }	
		\label{tab:models}
	\begin{ruledtabular}
			\begin{tabular}{c c c c c c c c c} 
				
				$T_{99}(s)$ & ${\mathcal E}_{\barnue} (10^{51} erg)$ & $\langle E_{\barnue}\rangle (MeV)$  & type & CR & CRM ($M_{\odot}$) & DM ($M_{\odot}$) & Model& Ref\\ [0.5ex] 
				\hline\hline
				0.58 & 4.4 & 18 & BNS & HMNS &3  &0.03 &Hinf &J.Lippuner(2017) \\ 
				
				0.40 & 2.0 & 16.5 & BNS&BH &3 &0.03 &B090  &  \\
				
				0.30 & 1.8 & 15.4 & NSBH&BH & 8.1& 0.1&BF15  &   \\
				\hline
				0.10 & 1.0 & 17.8 & BNS&BH  &3 &0.03 &M3A8m03a5 &O.Just(2015) \\
				
				0.27 & 11.2 & 16 & NSBH &BH & 6&0.3  & M6A8m3a5 &\\ [1ex] 
				\hline	
				0.99&14  & 10  &  BNS& HMNS & 2.7 &0.2  & DD2-1351350-On-H~~ & S.Fujibayashi(2017)
				\\
				\hline	
				0.58&40  & 20  &BNS  &HMNS  & 3 &   &M  &Y.Sekiguchi(2011)\\
				\hline	
				0.08&19.8& 24&NSBH &BH &6.6  & 0.49 &A5 & H.T.Janka (1999)
				\\
				
				0.16&23.2& 28& NSBH& BH&  11.6& 0.47 &B10 & \\
			
		\end{tabular}
	\end{ruledtabular}
	\end{table*}	

\begin{itemize}

	\item \underline{\emph{ Lippuner et al.}} \cite{Lippuner:2017bfm}, computed a long term ($\sim 10$ s simulation time) evolution of the accretion disk surrounding a HMNS of  variable lifetime, including a stable one (infinite lifetime, model Hinf), which we use here. 
	In this two-dimensional simulation, the HMNS (of mass $\sim$3 $M_\odot$) is approximated by using a reflecting inner boundary condition \cite{Metzger:2014ila}, and a parameterized  isotropic outward neutrino flux is imposed on the boundary.  The neutrino luminosity has a time dependence of $\approx t^{-\frac{1}{2}}$, which is general of the cooling of a proto-neutron star (see, e.g., \cite{Pons:1998mm}).
	When the HMNS collapses into a black hole, the boundary is changed from reflecting to absorbing, and the  \n\ emission is set to zero.
	Lippuner et al. also produced additional models with spinning black holes at the center; among these, we use models B090 and BF15. 
	
	\item \underline{\emph{Sekiguchi, Kiuchi, Kyutoku and Shibata}}  \cite{Sekiguchi:2011zd}, simulated \nsns\ \mgs, using the  Shen et al. \cite{Shen:1998gq} equation of state, and including effects of  general relativity and neutrino cooling. A long-lived HMNS is formed after the merger. Among their results, we use the one for a system of equal-mass neutron stars with an intermediate mass value,  $M_{NS}=1.5~ M_{\odot}$ (``middle" model, or M). 
	Because of the continuous collision between HMNS and spiral arms, the outer region of the HMNS is heated up, and, as a result, \n\ emission is strong \cite{Rosswog:2003rv,Dessart:2008zd}. Sekiguchi et al. found that a HMNS with mass $\lesssim 3 M_{\odot}$ has a life time much longer than its dynamical time scale ($t_{dyn}\sim$10 ms) and determined by the neutrino cooling process, with time-scale of the order of seconds. 
	The simulations by Sekiguchi et al. end at about 20 ms after the merger,  not long enough to estimate the total energy emitted in $\barnue$, which is needed here. 
	To circumvent this difficulty, we used the similarity between model M by Sekiguchi et al. and model  Hinf by Lippuner et al., which share similar HMNS mass and neutrino emission time scale. Considering the generality of the luminosity time dependence found in Lippuner et al. (see above), we used it to extend the results of Sekiguchi et al beyond 20 ms (with the appropriate normalization of the luminosity to match the numerical result at 20 ms, $L_{\barnue}\approx 2.3\times 10^{53} \,erg \,s^{-1}$).

	\item \underline{\emph{Fujibayashi, Sekiguchi, Kiuchi and Shibata}} \cite{Fujibayashi:2017xsz},  performed
	long term, two dimensional simulations of the evolution of the remnant of a \nsns\ \mgs\ -- with general relativity effects -- using initial conditions taken from a three-dimensional, numerical relativity simulation\cite{Sekiguchi:2015dma}.
	In their fiducial model (called DD2-135135-On-H, see \cite{Fujibayashi:2017xsz} for its detailed meaning)
	the $\barnue$ luminosity is  $L_{\barnue} \sim 10^{53} \,erg\,s^{-1}$ at the beginning of simulation, and decreases rapidly over $\sim$100 ms, setting to a nearly constant value of  $L_{\barnue} \sim 10^{52}\, erg\, s^{-1}$ after 300 ms. The simulation ends at 400 ms. To estimate the total energy in $\barnue$, we extrapolate the neutrino luminosity up to $t\sim 1$ s, similarly to the approach used by Kyutoku and Kashiyama  \cite{Kyutoku:2017wnb}.

	\item \underline{\emph{ Just, Bauswein, Pulpillo, Goriely and Janka}} \cite{Just:2014fka}, studied \nsbh\ and \nsns\ \mg\ resulting in BH-torus systems, using a 
	relativistic smooth-particle-hydrodynamic code \cite{Oechslin:2001km}, with Newtonian hydrodynamics, and including viscosity effects. 
	They found that the masses of black hole and of the torus are important parameters, strongly influencing both the neutrino-driven wind and the  neutrino luminosities.
	The neutrino energy loss rates can reach $L_\nu \sim few~ 10^{52}\,erg\,s^{-1}$, and even exceed  $L_\nu \sim 10^{53}\,erg\,s^{-1}$ for each of $\nu_e$ and $\barnue$ for a few hundreds of ms. For more detail on these models, we refer to  Table  \ref{tab:models} (see also Table 2 of \cite{Just:2014fka}). 

	\item \underline{\emph{Janka, Eberl, Ruffert and Fryer}} \cite{Janka:1999qu},  performed Newtonian hydrodynamic simulations of \nsbh\ and \nsns\ mergers using a relativistic EoS \cite{Lattimer:1991nc}. In their simulations, the maximum temperature $T_{max}$ of the accretion flow can reach several tens of MeV,
	resulting in efficient \n\ emission 
	at densities  $\rho \sim 10^{10}-10^{13}\,{\rm g\,cm^{-3}}$.
	Total energies  ${\mathcal E}_\nu\approx 3\times 10^{52}\, erg$ are found for the \n\ emission. The parameters of their models are given in details in Table  \ref{tab:models} (see also Table 2 in \cite{Janka:1999qu}).  Note that the time duration of the $\barnue$ luminosity in the Janka et al. model is approximated by the authors to be the same as torus life, which is around $50-150$ ms. We assume this to be the time $T_{99}$ shown in the Table.
	
\end{itemize}

\subsection{\label{sub:flux}Cosmological merger neutrino flux}

For a given class of \mgs\ (\nsns\ or \nsbh), the  cosmological \n\ flux  depends on the volumetric merger rate (number of mergers per unit of comoving volume, per unit time), as a function of the redshift, $R(z)$.  
This quantity  depends on the core collapse supernova rate, which in turn tracks the star formation rate with its characteristic redshift  evolution like $\sim (1+z)^3$ (for $z\lesssim 1$). It also depends on the post-collapse evolution of the remnants, which is only partially understood; therefore uncertainties on the merger rate are large (see, e.g. \cite{Dominik:2013tma,Belczynski:2015tba,Chruslinska:2018ylm,Mapelli:2018wys} for overviews).

\begin{figure}[htbp]
	\centering
	\includegraphics[width=0.45\textwidth]{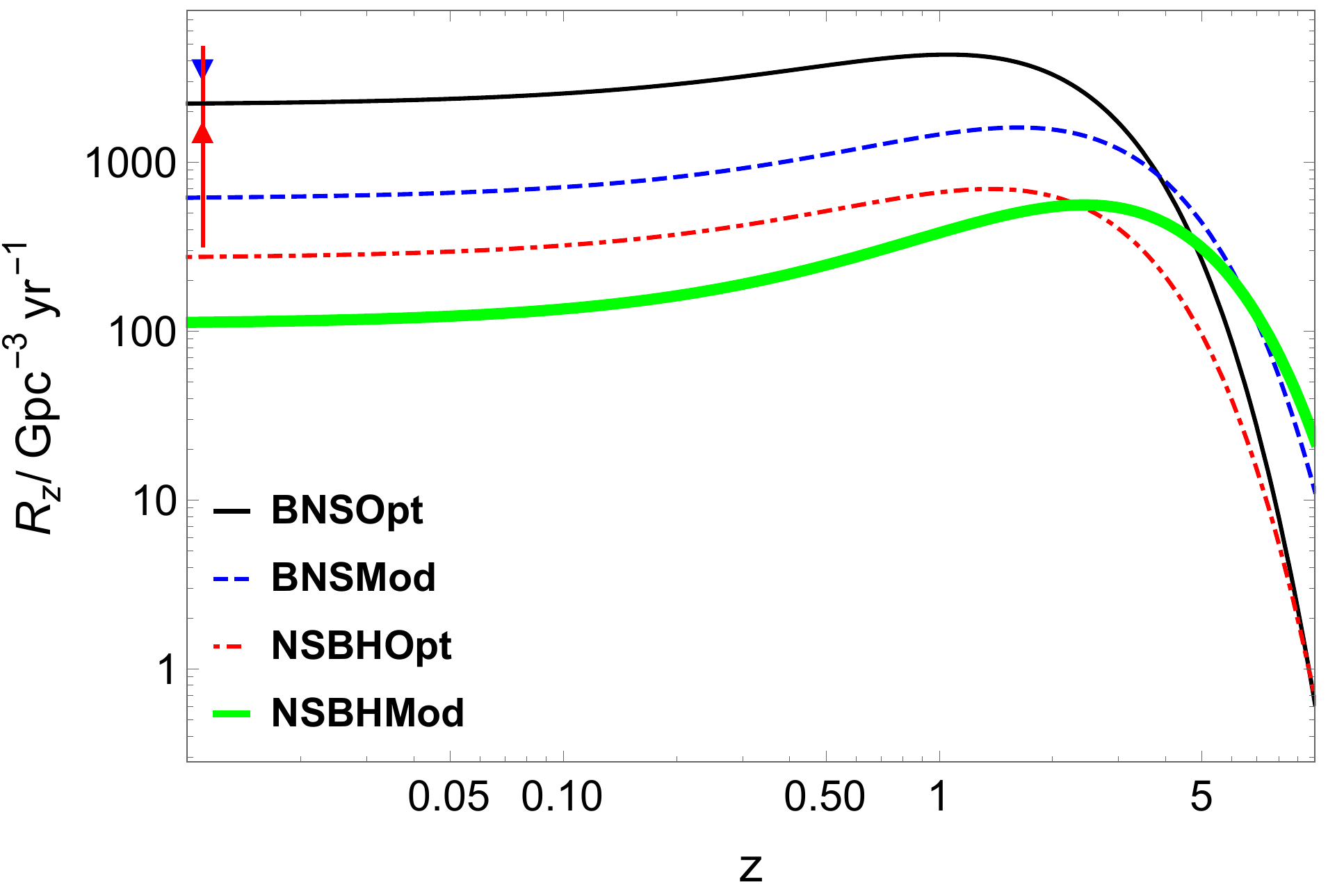}
\caption{ Volumetric merger rates, as a function of the redshift, for \nsns\ and \nsbh\  mergers. For each merger type, the upper (optimistic rate) and lower (moderate rate) curves are from Eldridge, Stanway and Tang \cite{Eldridge:2018nop}, and from  Mapelli and Giacobbo \cite{Mapelli:2018wys}, respectively (see legend).  The measurement/bound from \lgvg\ are shown for comparison. Red solid error bar  with red triangle: BNS merger rate inferred from GW170817 \cite{TheLIGOScientific:2017qsa}. Blue down triangle:  upper limit to the NSBH merger rate inferred from the LIGO O1 run \cite{Abbott:2016ymx}. The position of the \nsbh\ upper limit and the \nsns\ constraint along the horizontal axis is just for visualization purposes.}
\label{fig:mgRate}
\end{figure}

For each \mg\ type, we use two representative models for $R(z)$, whose predictions are consistent with the \lgvg\  observational bounds. The first, more optimistic (``optimistic" from here on), is from the work of Eldridge, Stanway and Tang \cite{Eldridge:2018nop}. Specifically, we use the results for the kick model there (see also \cite{Bray:2018nzk} for a description of the model itself) \footnote{In \cite{Eldridge:2018nop}, the redshift dependence of the rate for the kick model is not explicitly given. So, we constructed an approximate analytical form by taking an analytical expression used for other models in the same paper, with parameters adjusted to fit the tabulated rates (Table 2 in \cite{Eldridge:2018nop}) for the kick model.  }. 
The second model, more conservative (``moderate" from here on), is from  Mapelli and Giacobbo \cite{Mapelli:2018wys} (case labeled as $\alpha=5,~$ low $\sigma$ in \cite{Mapelli:2018wys}). Both models are illustrated in Fig. \ref{fig:mgRate} for \nsns\ and \nsbh\ mergers.

To estimate the neutrino flux from mergers in a detector, one needs to use the comoving volume enclosed between redshift $z$ and $z+dz$, which can be written in terms of the comoving distance, $D_c$, as follows:
\begin{eqnarray}
\frac{dV}{dz}&= &\frac{4\pi D^2_c c}{H(z)}~,\\
D_c&=	&\int_{0}^{z}\frac{c}{H_0\sqrt{\Omega_m(1+\tilde{z})^3+\Omega_\Lambda}} d\tilde{z}~. 
\end{eqnarray}

The rate of mergers with $z<z_{max}$, in the frame on an observer on Earth, is then given by: 
\begin{equation}	N(z_{max})=\int_{0}^{z_{max}}\frac{R(z)dV}{(1+z)dz} dz~. 
\label{equ:nummerg}
\end{equation}
Here, $\Omega_{ m}=0.3$ and $\Omega_\Lambda=0.7$ are the fractions of the cosmic energy density in matter and dark energy respectively; $c$ is the speed of light and $H_0$ is the Hubble constant. 
We assume the $F_{\barnue}(E)$ follows Firmi-Dirac distribution with zero chemical potential. Therefore, the (time-integrated) number of $\barnue$ emitted per unit energy by an individual merger, $F_{\barnue}(E)$, is: 
\begin{equation}	F_{\barnue}(E)=\frac{{\mathcal {E}}_{\barnue}}{\langle E_{\barnue}\rangle}\frac{2}{3T^3\zeta(3)}\frac{E^2}{e^{E/T}+1}, 
\label{equ:mergerflux}
\end{equation}
where ${\mathcal{ E}}_{\barnue}$ is the total  neutrino energy emission, $\langle E_{\barnue}\rangle$ is the mean neutrino energy, $T$ is the neutrino temperature, and $2/3T^3\zeta(3)$ serves as a normalization factor of the Fermi-Dirac distribution.
Given $F_{\barnue}$,the total, time averaged flux (differential in energy, surface and time) at Earth due to all mergers with $z<z_{max}$ can be expressed as:
\be
\Phi_{\barnue}(E)= f_{osc}\frac{c}{H_0}\int_0^{z_{ max}} R(z)
F_{\barnue}(E^\prime)
\frac{{d} z}
{\sqrt{\Omega_{ m}(1+z)^3+\Omega_\Lambda}}~ ,
\label{equ:flux}
\ee
where $E' = E(1+z)$, describes the redshift of the \n\ energy during propagation. The constant $f_{osc}$ is a  phenomenological factor accounting for the effect of flavor oscillations.  Detailed studies of \n\ oscillations in the post-merger environment for \nsns\ and \nsbh\ mergers have shown a complex pattern that depends strongly on the relative intensity of the different flavor fluxes and other parameters describing the compact object and the accretion disk. The $\barnue$ survival probability varies, roughly, between $0.5$ and 1 (corresponding to no oscillations), see, e.g.,  \cite{Malkus:2012ts, Malkus:2015mda,Zhu:2016mwa,Tian:2017xbr,Vlasenko:2018irq}. Therefore, here we set  $f_{osc}=1$. 

\section{Detectability}
\label{sec:rates}

\subsection{The method; expected backgrounds}
\label{sub:method}

We consider a near-future scenario where: (i) a \n\ detector of mass ${\mathcal O(100)}$ kt or higher exists, with good timing resolution and low energy threshold, like the upcoming HyperKamiokande; and (ii) one or more next generation \gw\ detectors are available, with the capability to observe mergers  with $\sim 100\%$ efficiency, up to a redshift $z_{GW}\gtrsim 1$.   We also assume that the \gw\ data will allow to establish the time of the merger and its distance/redshift with good precision (e.g., tens of per cent)\cite{Safarzadeh:2019zif}.  

In this scenario, a realistic detection method would be the one proposed by Kyutoku and Kashiyama \cite{Kyutoku:2017wnb}, which is generalized here to account for non-zero redshift. The method consists of considering the $N$ mergers that are observed in \gw\ over a long period of time, $\Delta T$, with their merger times $t_i$ ($i=1,2,...,N$), and redshifts $z_i$, and restricting the \n\ data analysis to time windows of width $\Delta t_i \sim T_{99}(1+z_i)$ after each merger, where $T_{99}$ is the time duration of neutrino burst within which  $99\%$ of $\nu_{\bar{e}}$ energy  has  emitted. Here the factor $(1+z_i)$ ensures that each time windows is appropriately adjusted for the cosmological time dilation. Thus, the number of \n\ events collected in the detector over $\Delta T$ is:
\beq
N_s = \sum^{N}_{i=1} n_s(t_i) \Delta t_i 
\label{equ:signal} 
\eeq
 where, $n_s$ is the (average) \n\ event rate per merger. For the same observation time, the expected number of background events in the detector has an expression analogous to \equ{signal}, where the rate of background events, $n_b(t)$, can be approximated as constant in time: $N_b \simeq n_b \sum^{N}_{i=1} \Delta t_i$.   The effect of restricting to the effective observation time, $\Delta T_{eff}=\sum^{N}_{i=1} \Delta t_i$ results in a strong enhancement of the signal-to-background ratio \cite{Kyutoku:2017wnb}. 

For simplicity, we neglect the possible corrections due to the \n\ mass and to gravitational lensing on the expected time window $\Delta t_i$ \cite{Stodolsky:1999kc,Baker:2016reh} (see also discussion in \cite{Kyutoku:2017wnb}), as well as effects of possible physics beyond the Standard Model of particle physics. We also neglect differences in the \n\ emission between individual mergers, thus making the somewhat simplistic assumption that all mergers of the same class are identical.  We estimate that the corrections due to these neglected effects would be sub-dominant compared to the very large uncertainties on the \n\ emission models and merger rates. 

Let us consider a water Cherenkov detector of the next generation,
and its main channel of detection, inverse beta decay: $\barnue + p \rightarrow n + e^+$. One can approximate the sum for $N_s$ as an integral of the diffuse flux over the observation time $\Delta T$:
\beq
N_s  \simeq  N_{t} \Delta T \int_{E_{th}}^{E_{max}} \eta(E) \Phi_{\bar e}(E) \sigma(E) dE  ~. 
\label{equ:signal2}
\eeq

In this approximation, it is assumed that the choice of the time windows, $\Delta t_i$, allows to capture the entire post-merger \n\ flux, i.e., $\Delta t_i/(1+z_i)>T_{99}$. 
In this work, a rather conservative time window, $\Delta t_i/(1+z_i)=1$ s (to be compared with $T_{99}$ for various models, see Table  \ref{tab:models}), will be used in all calculations. 
Here $\sigma(E)$ is the detection cross section and $N_t$ is the number of protons in the detector;  $\eta(E)$  is the detector efficiency. 
The interval $E= E_{th} - E_{max}$ is a suitable energy window, determined by the need to minimize the background, see \Sec\ \ref{subsub:rates}. 
\begin{figure}[htbp]
	\centering
	\includegraphics[width=0.45\textwidth]{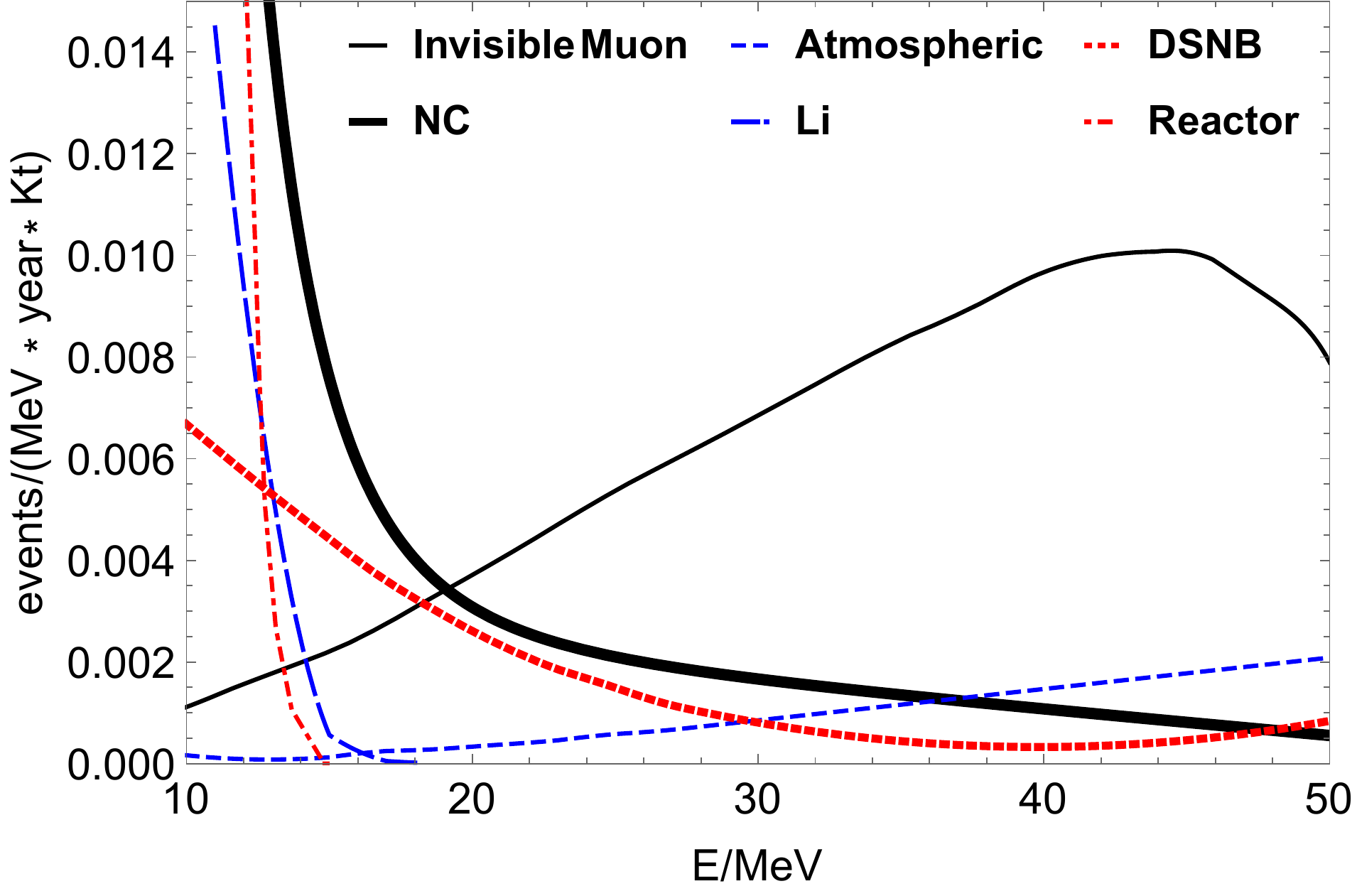}
	\includegraphics[width=0.45\textwidth]{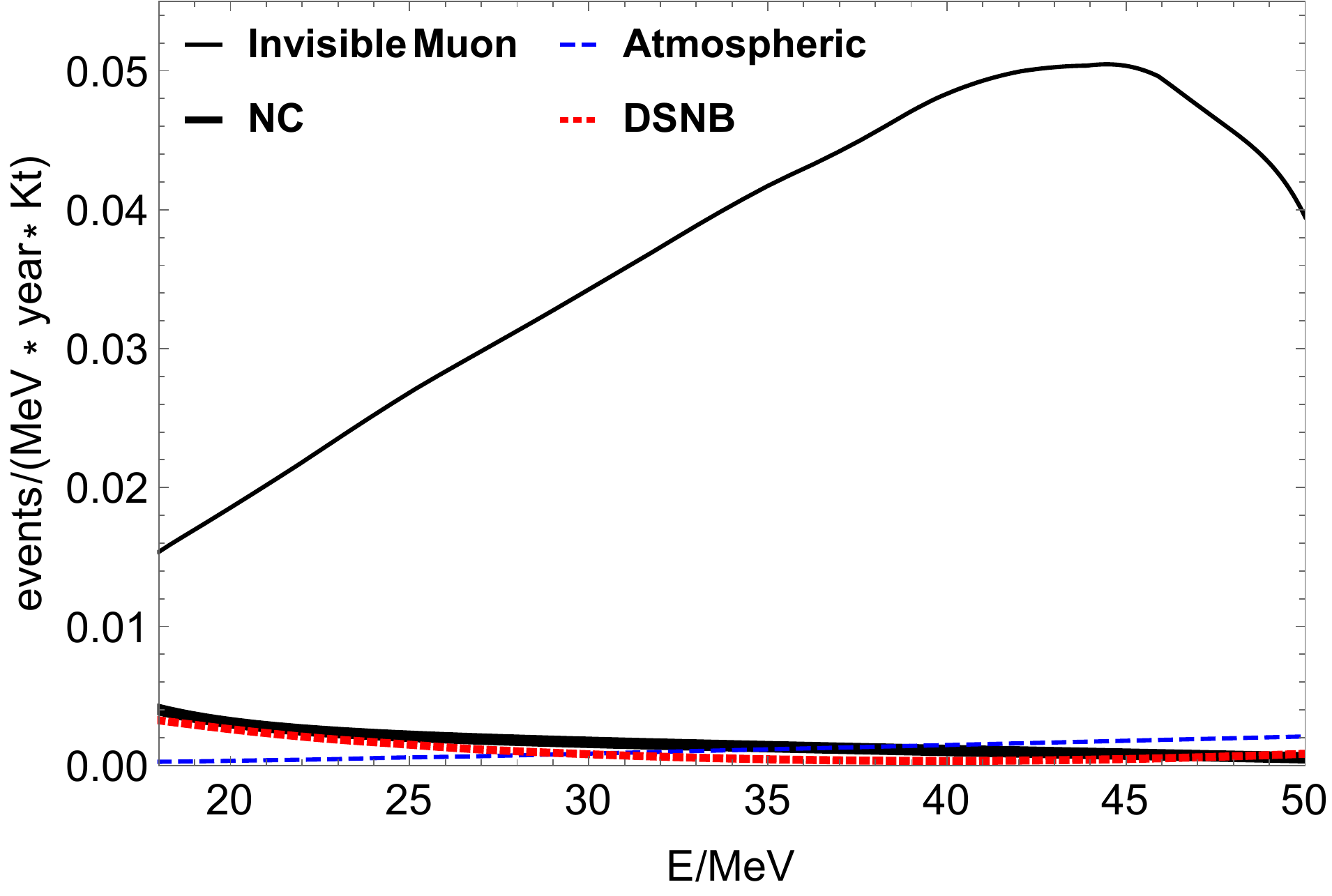} 
	\caption{ Rates of various types of backgrounds, differential in visible energy, for a water Cherenkov detector with (upper panel) and without (lower panel) Gadolinium, from \cite{phdthesis1,Abe:2018uyc}. The backgrounds presented here are: invisible muon background (Invisible Muon), atmospheric charged current background (Atmospheric), atmospheric neutral current background (NC), Lithium background (Li), background due to the diffuse supernovae neutrino flux (DSNB), and background due to reactor neutrinos at SuperKamiokande location (Reactor).   
	}
	\label{fig:bckg}
\end{figure}
Let us now discuss the experimental background. We consider two different detector configurations, the first with pure water, and the second with water with the addition of Gadolinium (Gd) \cite{Beacom:2003nk}, for better background reduction. The option with Gd is being realized, for the first time, in the upgraded SuperKamiokande (SuperK-Gd; see, e.g., \cite{Marti-Magro:2017pzo}), and is envisioned as a possibility for a second phase of HyperKamiokande \cite{Abe:2018uyc}.  
For pure water, the region of visible energy $E_{vis}\lesssim 18$ MeV is unaccessbile due to the overwhelming spallation background. At higher energy,  the background is dominated by invisible muons (\figu{bckg}, bottom panel): these are sub-Cherekov atmospheric muons that become visible in the detector only when they decay, thus mimicking inverse beta decay. We take  $\eta \simeq 0.9$ as a realistic efficiency \cite{Hirata:1988ad,Kyutoku:2017wnb,Abe:2011ts}. 

For water with Gd, the background due to spallation can be effectively reduced, with only a small residual contribution due to the production of a short-lived Lithium isotope (${\rm^9Li}$).  Therefore, the window of sensitivity extends down to $E_{vis}\sim 11$ MeV, below which events due to reactor $\barnue$ dominate. With Gd, the invisible muon background is suppressed by a factor of $\sim 5$ \cite{Beacom:2003nk}, and the total background at $E_{vis} \gtrsim 11$ MeV has comparable contributions from different sources. 
Specifically, besides invisible muons, one should include inverse beta decay events due to the diffuse supernova \n\ background and to atmospheric $\barnue$, as well as events due to neutral current scattering of atmospheric neutrinos. 
For illustration of these backgrounds, we refer to the specific set of background cuts that was discussed for the diffuse supernova \n\ background at SuperK-Gd and HyperKamiokande \cite{phdthesis1,Abe:2018uyc}. These result in a signal efficiency $\eta \simeq 0.67$, and produce the residual background spectrum shown in \figu{bckg}, top panel. 


\subsection{Results}
\label{sub:results}

\subsubsection{Event rates and physics potential}
\label{subsub:rates}

In this section, results will be shown for the signal and background event rates. Unless stated otherwise, they refer to a detector exposure $T_{exp}=100~{\rm Mt ~yr}$, corresponding, e.g. to a 1 Mt (10 Mt) mass detector running for a century (decade).  While such exposure does not seem entirely realistic for HyperKamiokande as currently planned (total fiducial mass of 374 kt, see \cite{Abe:2018uyc}), it is technlogically feasible, and within the realm of possibility for the next-to-next generation water Cherenkov detectors. 
For illustration, here several results will be shown for \n\ parameters (${\mathcal E}_{\barnue}$ and $\langle E_{\barnue}\rangle$) of the specific models in Table \ref{tab:models}. Considering the uncertainties that affect numerical simulations (see Sec. \ref{sub:models}), results that are intermediate between models should also be considered possible.   
%
%

\begin{figure}[htbp]
	\centering
	\includegraphics[width=0.45\textwidth]{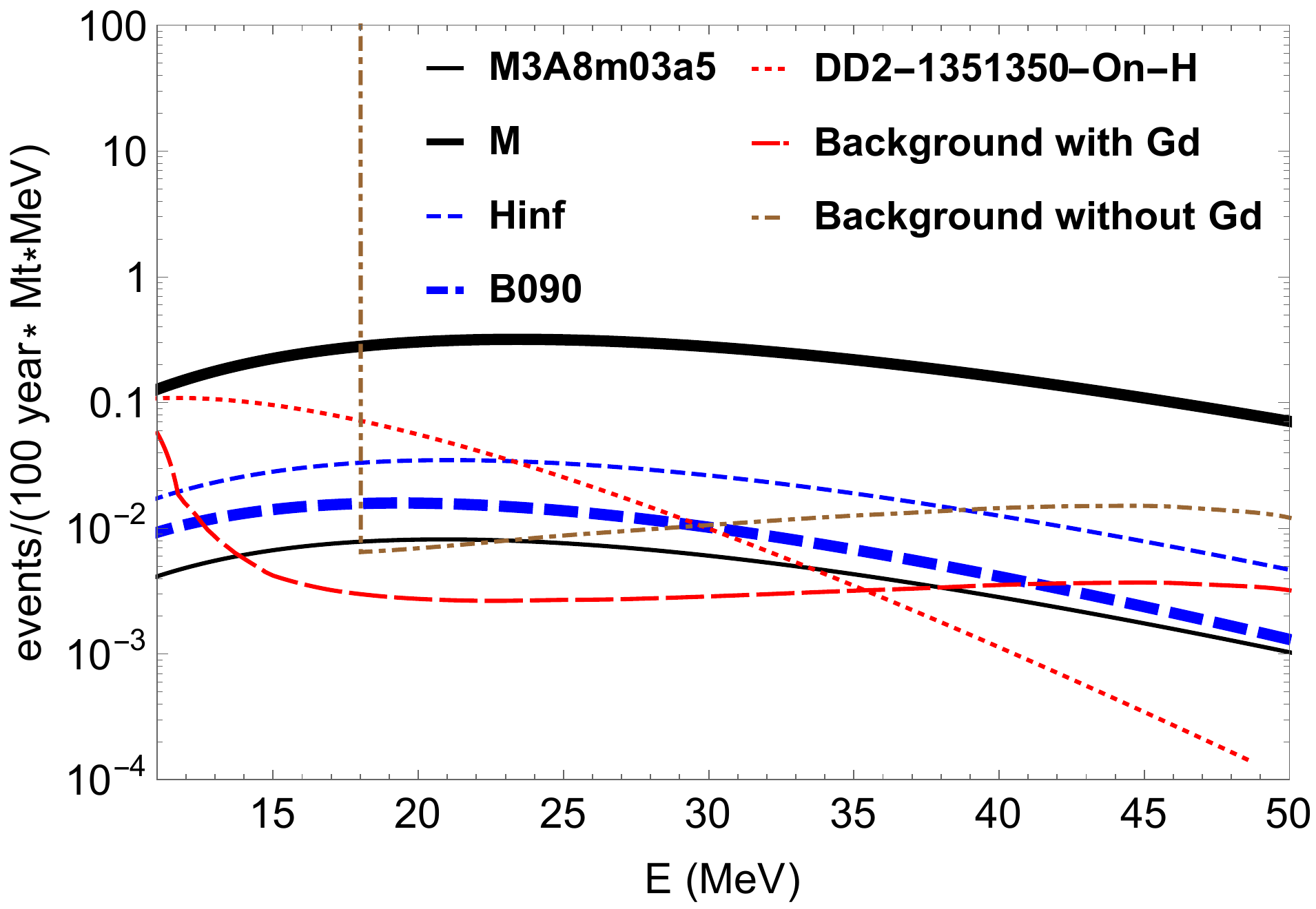} 
	\includegraphics[width=0.45\textwidth]{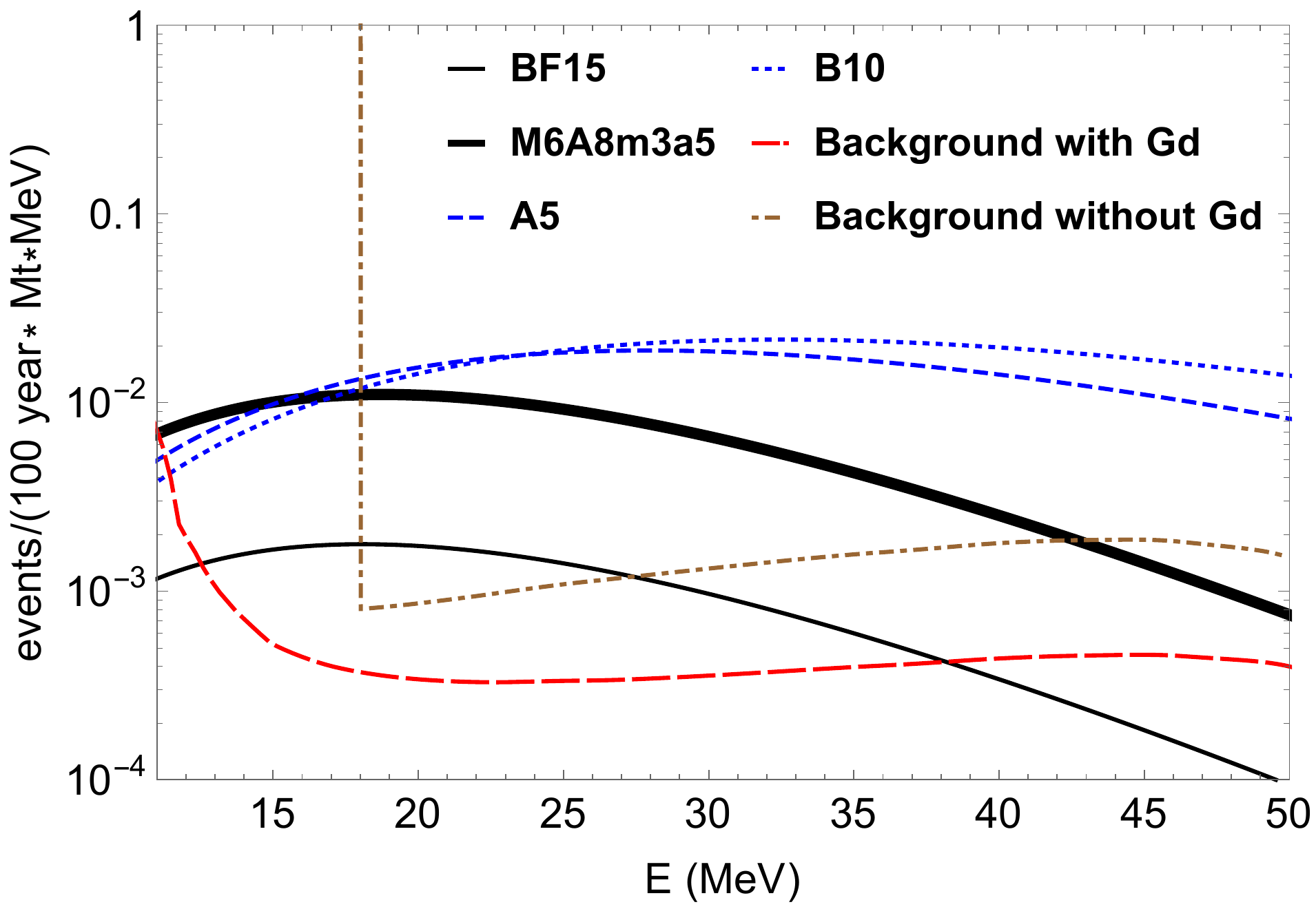} 
	\caption{ The spectrum of  background events (for configuration with and without Gd) and of signal events for different models of \nsns\ (upper panel) and \nsbh\ (lower panel) mergers (see Table  \ref{tab:models}) for the redshift bin $z=0 - 0.05$.  The optimistic merger rates (see Fig. \ref{fig:mgRate}) have been used here. For definiteness, only signals for water with Gd (detection efficiency $\eta\simeq 0.67$) are shown; the case of pure water is nearly identical, differing only by the value of the efficiency ($\eta \simeq 0.9$), see text.}
	\label{fig:enwindow}
\end{figure}
To assess the detectability prospects, let us first examine the energy spectra of signal (in terms of observed positron energy, $E_e$) and background events, 
and establish the interval of energy where the signal exceeds the background. 
The spectra are shown in \figu{enwindow} for  mergers of both types (\nsns\ and \nsbh), with redshift in the interval $z=0 - 0.05$.  For water with Gd, it appears that the energy window of signal dominance is $E_{e} \sim 14- 34$ MeV for the most conservative models. It extends down to $E_{e} \sim 11$ MeV and beyond 50 MeV for the most  luminous mergers of either type. For pure water, the energy window is bound from below by the threshold due to spallation (see Sec. \ref{fig:bckg}); its upper edge is as low as $\sim 25$ MeV or as high as 50 MeV or beyond depending on the signal flux model. 

In principle, the  energy window depends on the redshift of the merger of interest, because of the effect of the redshift of energy on the signal spectrum. We explored the possibility of using a $z-$dependent window to further reduce backgrounds, and found that it provides some advantage for the models with lower \n\ emission, but is slightly detrimental for the most optimistic models, where the energy window extends beyond 50 MeV for a wide range of $z$.   

For definiteness, we settle on a moderately conservative and $z-$independent energy window, $E_{e}= 18-50$ MeV ($E_{e}= 11-50$ MeV) for pure water (water with Gd)\footnote{The decision to limit  the analysis to $E_{e}<50$ MeV is in part motivated by the fact that the background is poorly known at higher energy, where contributions from the inelastic scattering of high energy atmospheric \ns\ becomes increasingly important.}. 

\begin{figure}[htbp]
	\centering
	\includegraphics[width=0.45\textwidth]{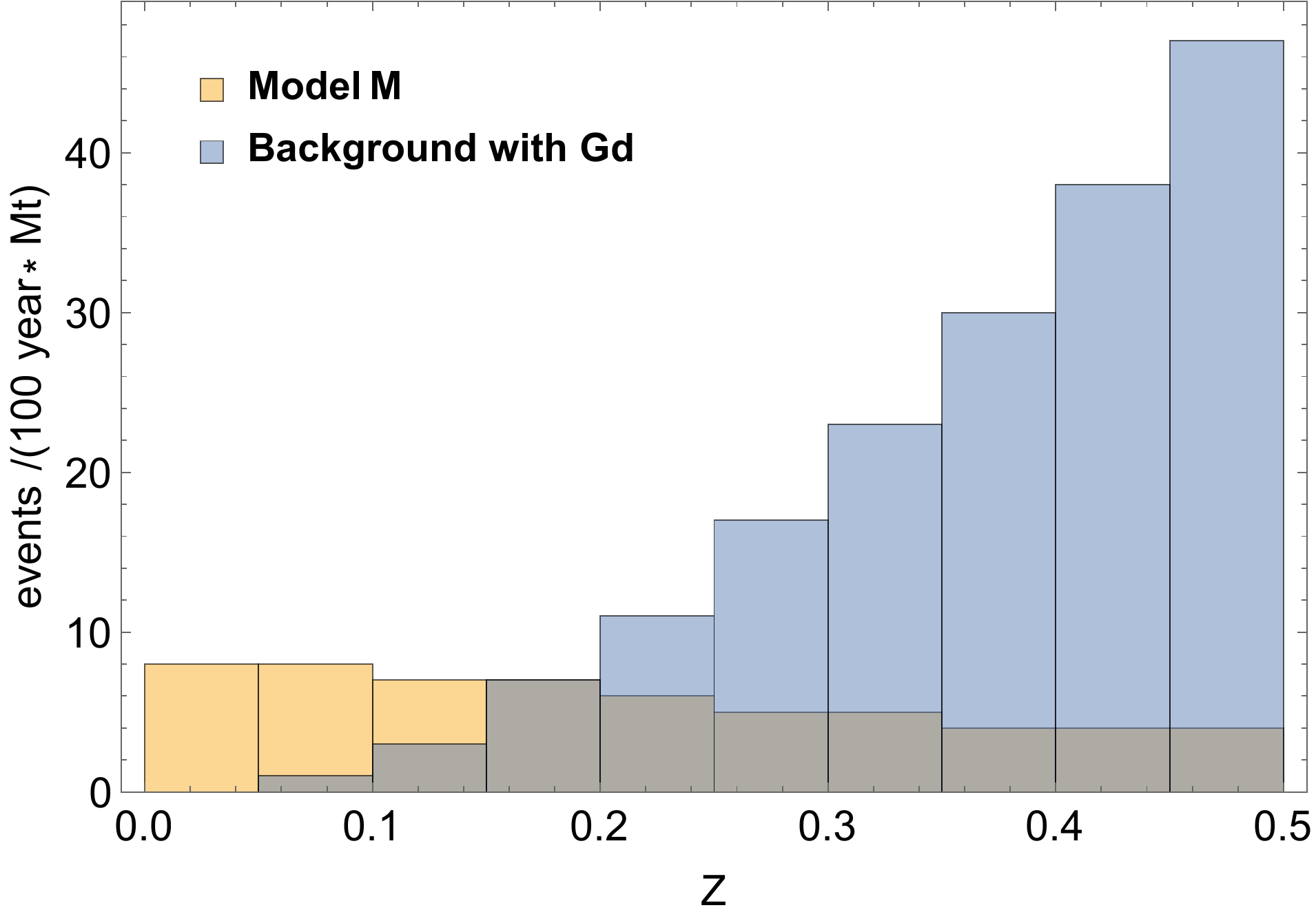} 
	\vskip 0.3truecm
	\includegraphics[width=0.45\textwidth]{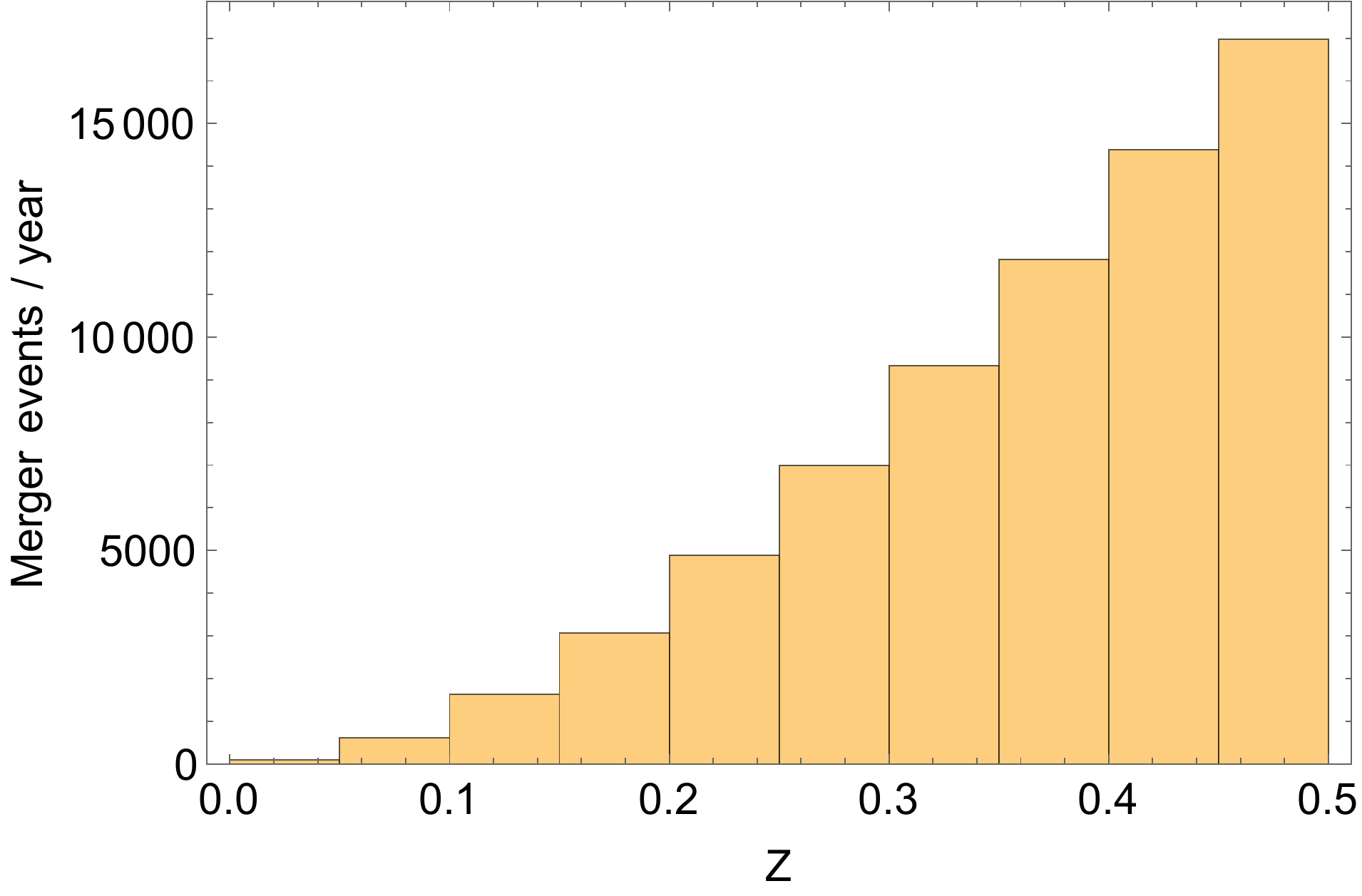} 
	\caption{{\it Upper:} Number of signal and background events (for water with Gd) in bins of redshift of width $\Delta z=0.05$.  {\it Lower:} the corresponding number of mergers detected in GW per year.   The optimistic merger rate (see Fig. \ref{fig:mgRate}) and the \nsns\ merger model spectra by Sekiguchi et al. (model M, see Table  \ref{tab:models}) have been used here. 
	}
	\label{fig:zbins}
\end{figure}
\begin{figure*}[htp]
	\centering
\includegraphics[width=0.45\textwidth]{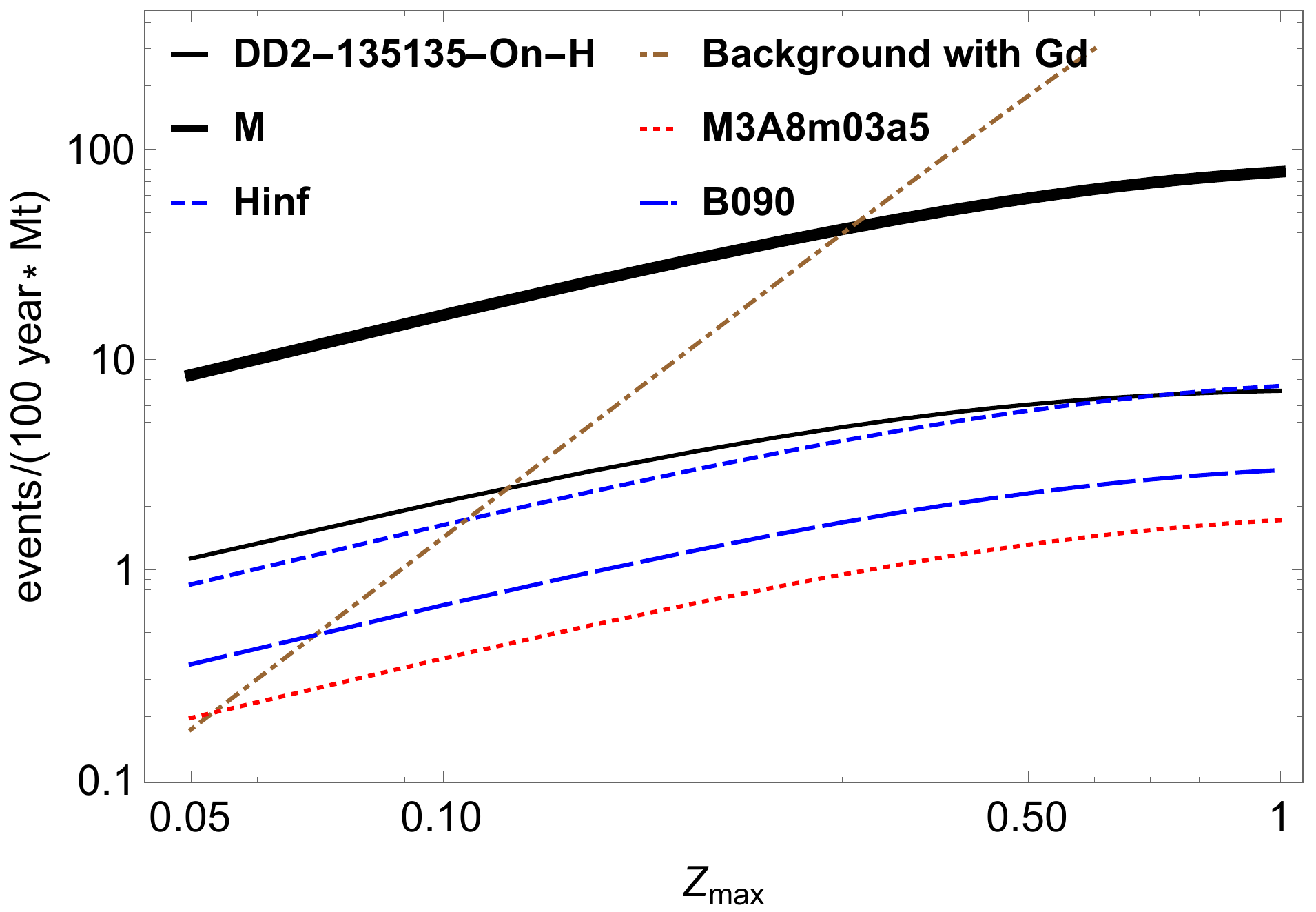}
	\includegraphics[width=0.45\textwidth]{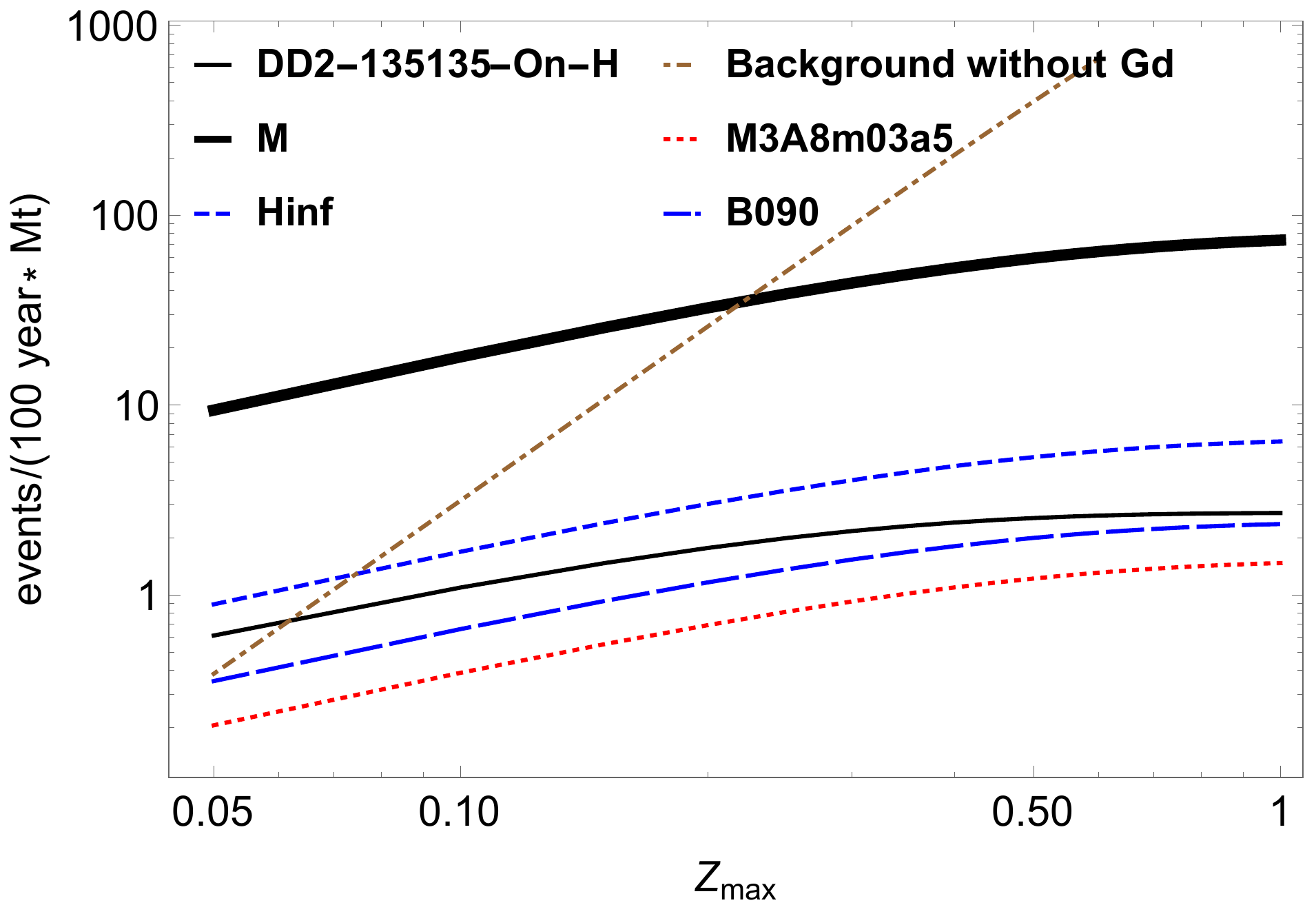} 
	\includegraphics[width=0.45\textwidth]{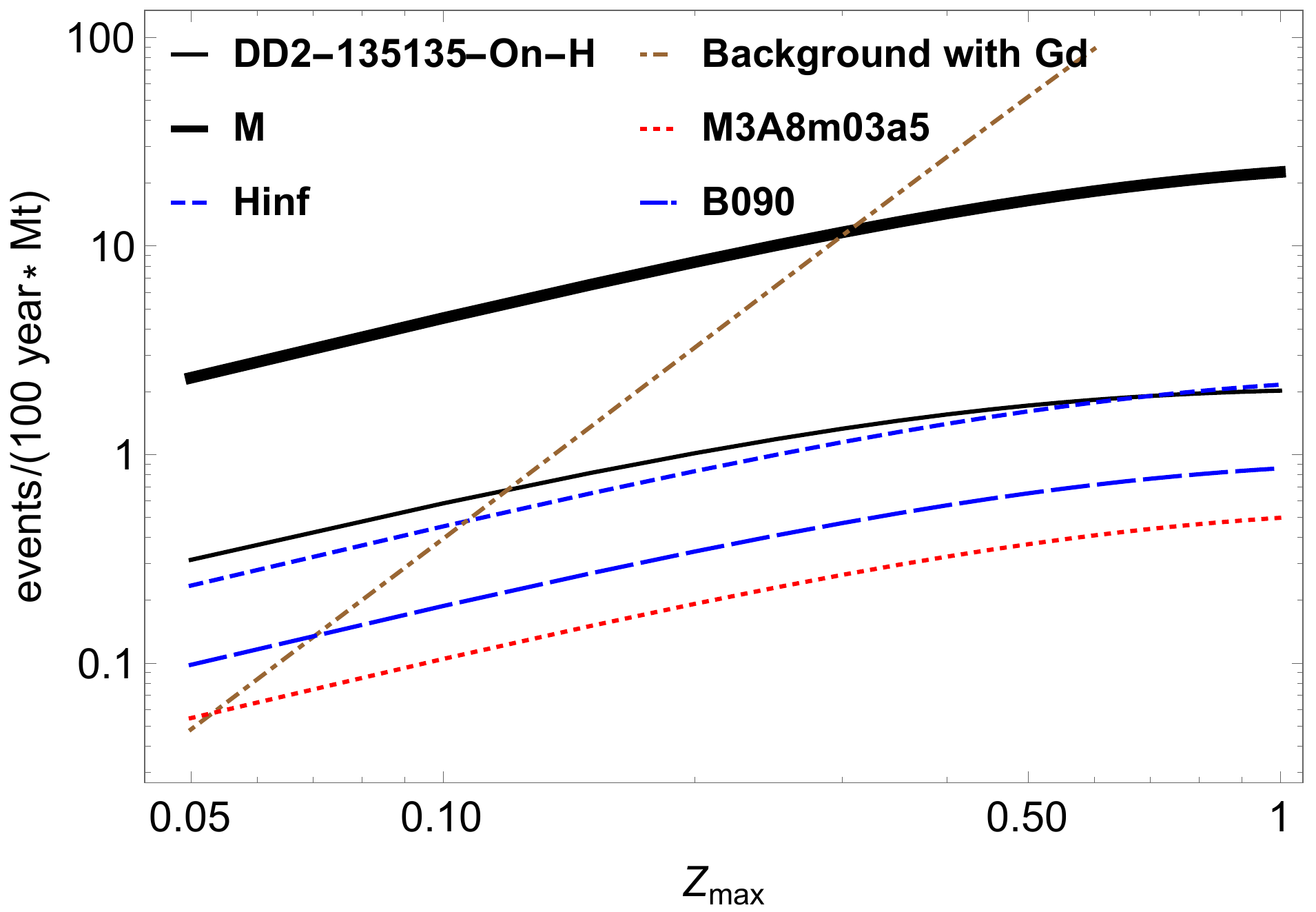} 
	\includegraphics[width=0.45\textwidth]{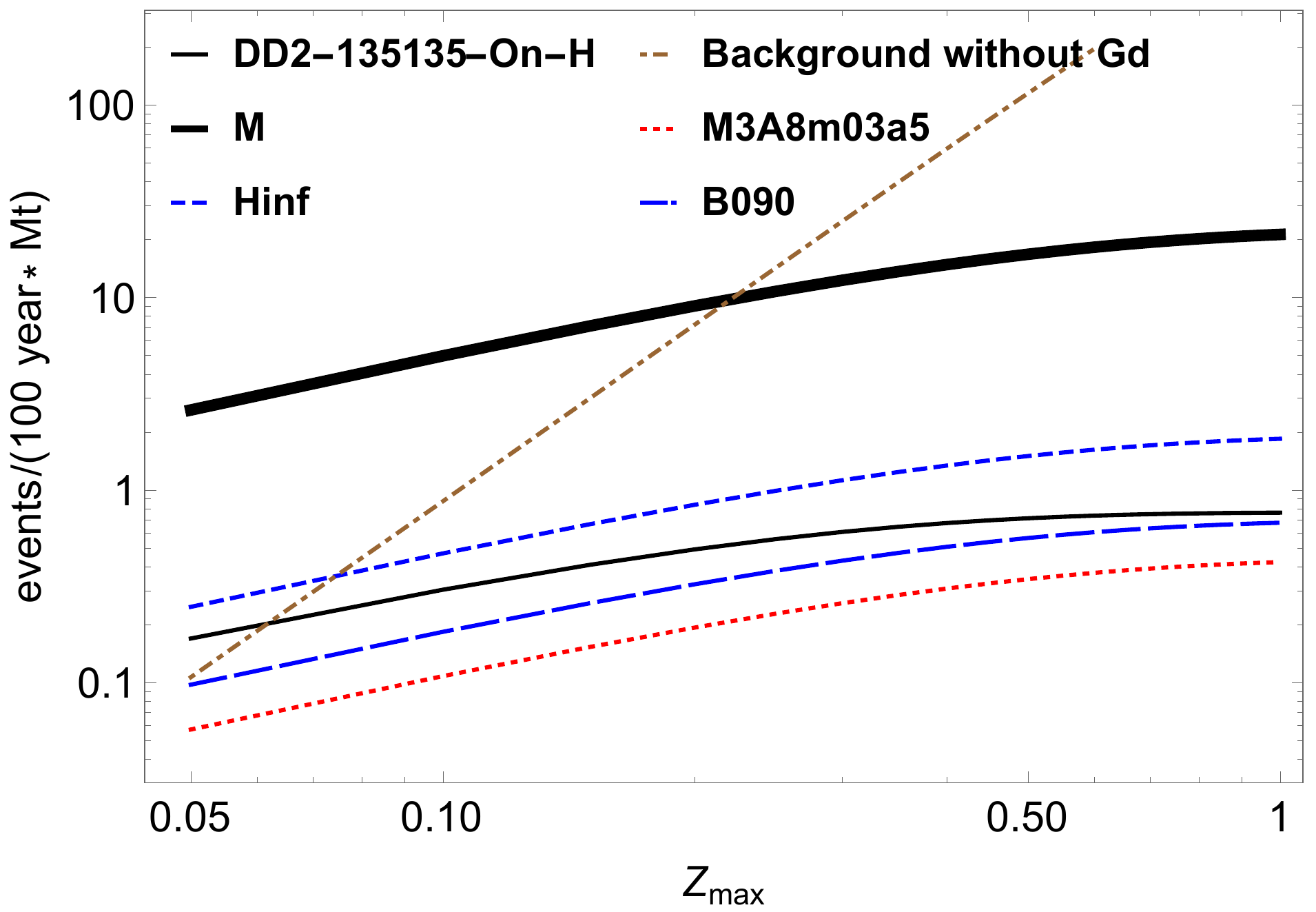} 
	\caption{Number of signal events for \nsns\ \mgs, and of background events in the energy window, for  a redshift window $z \in
		[ 0, z_{max}] $, as a function of $z_{max}$ (see \equ{signal2}). Results are shown for different \n\ emission models (see Table  \ref{tab:models}), for detector configuration with and without Gd, and for the optimistic (upper row) and conservative (lower row) merger rate. }
	\label{fig:ratesz}
\end{figure*}
\begin{figure*}[htp]
	\centering
	\includegraphics[width=0.45\textwidth]{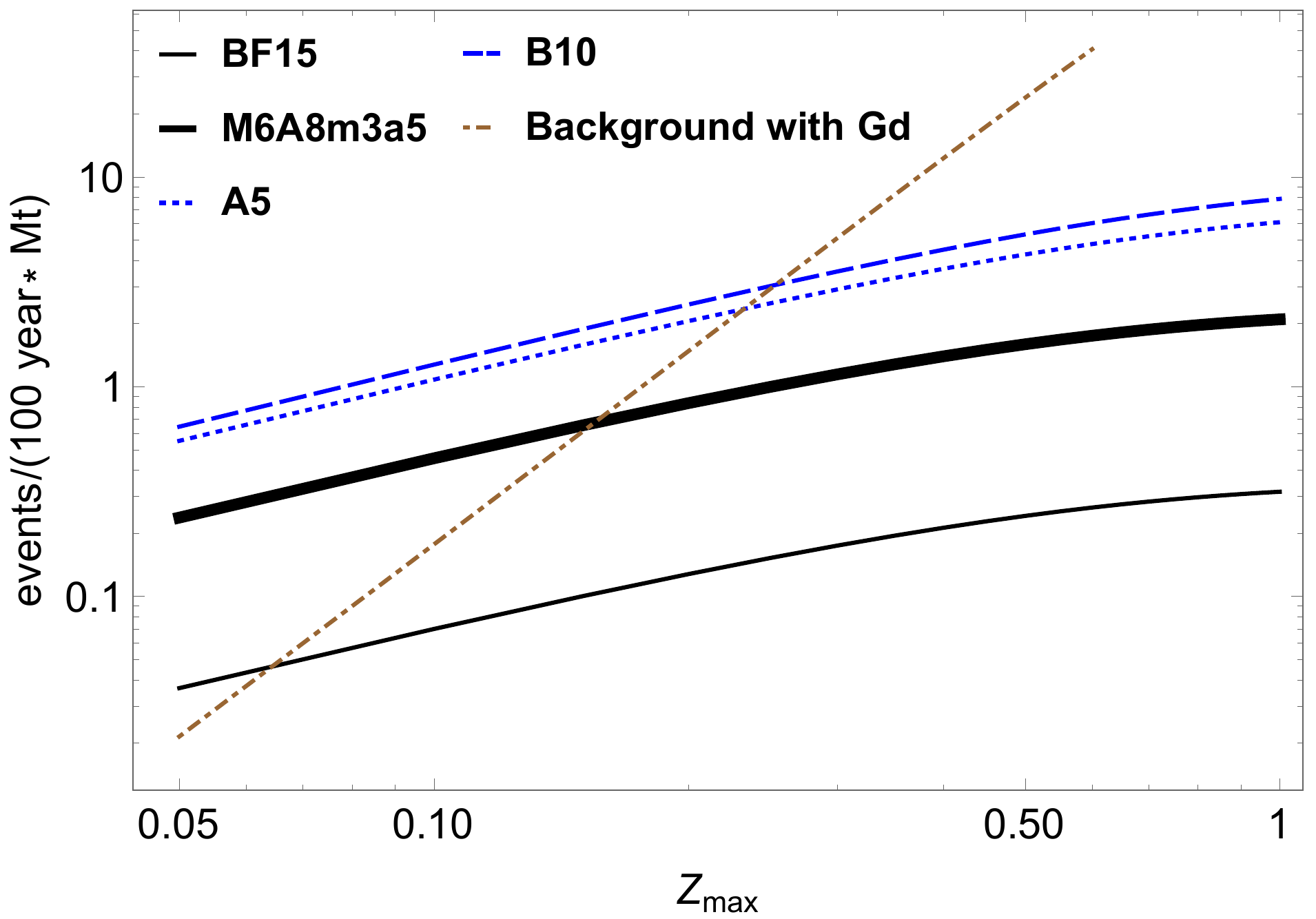} 
	\includegraphics[width=0.45\textwidth]{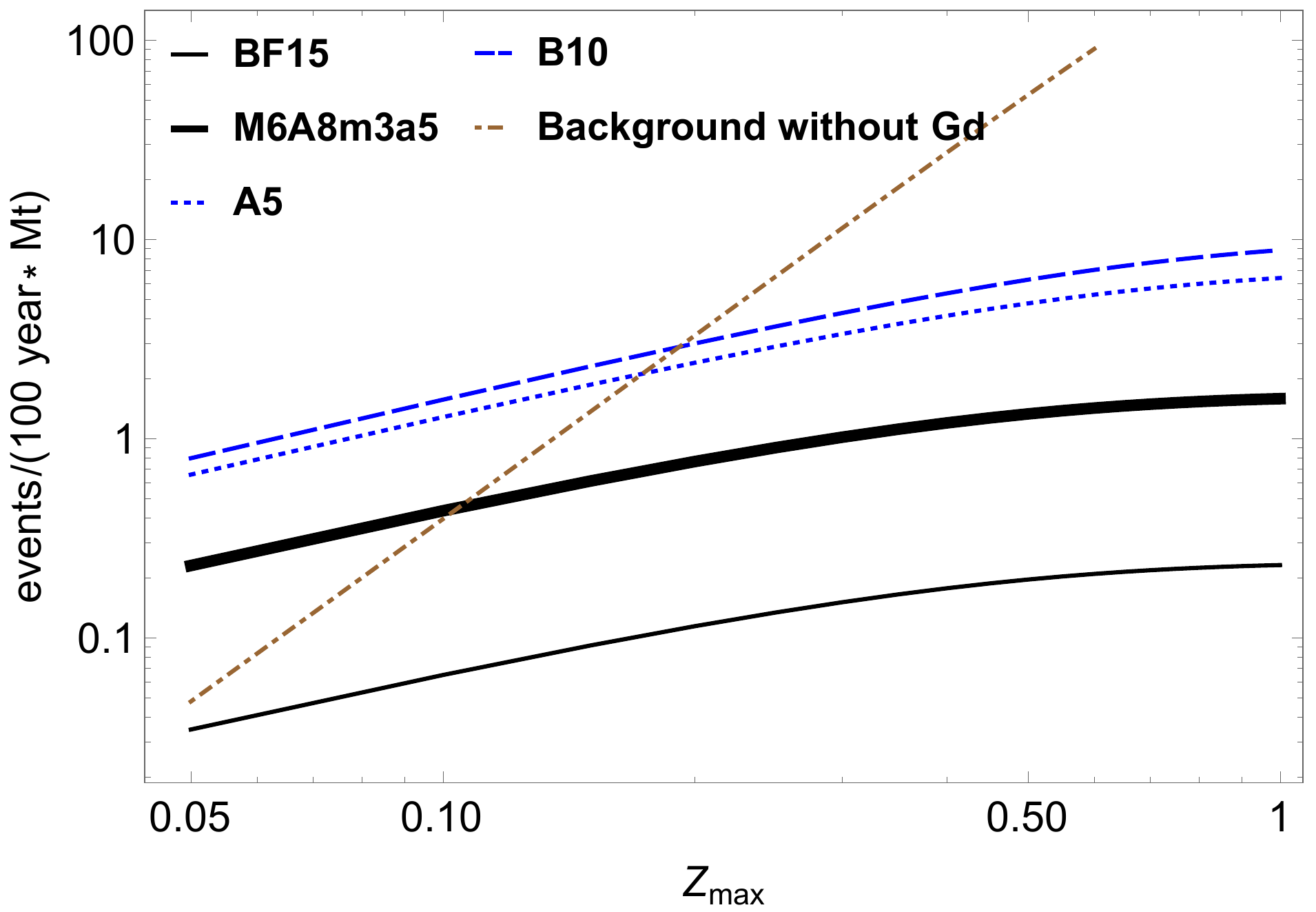} 
	\includegraphics[width=0.45\textwidth]{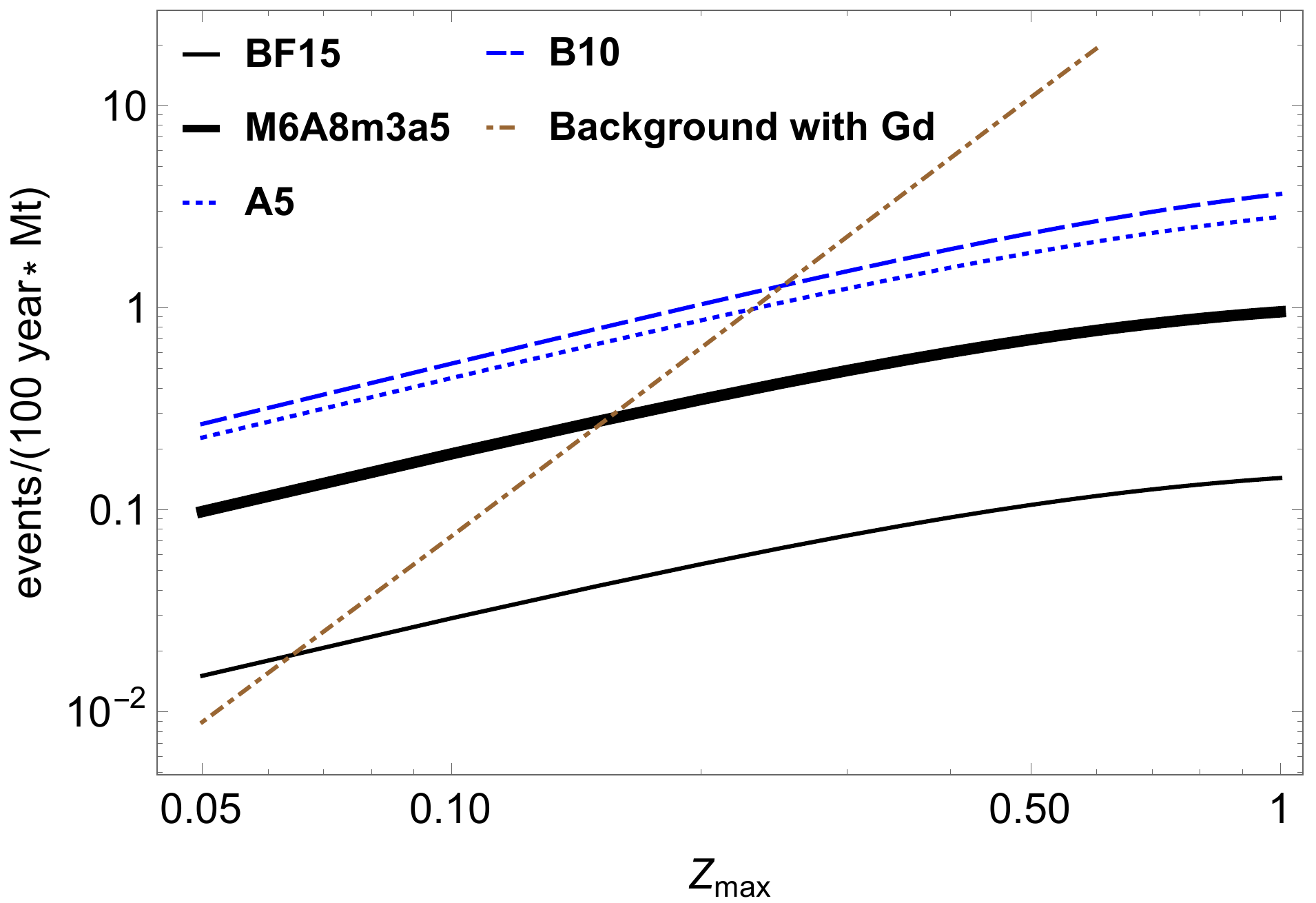} 
	\includegraphics[width=0.45\textwidth]{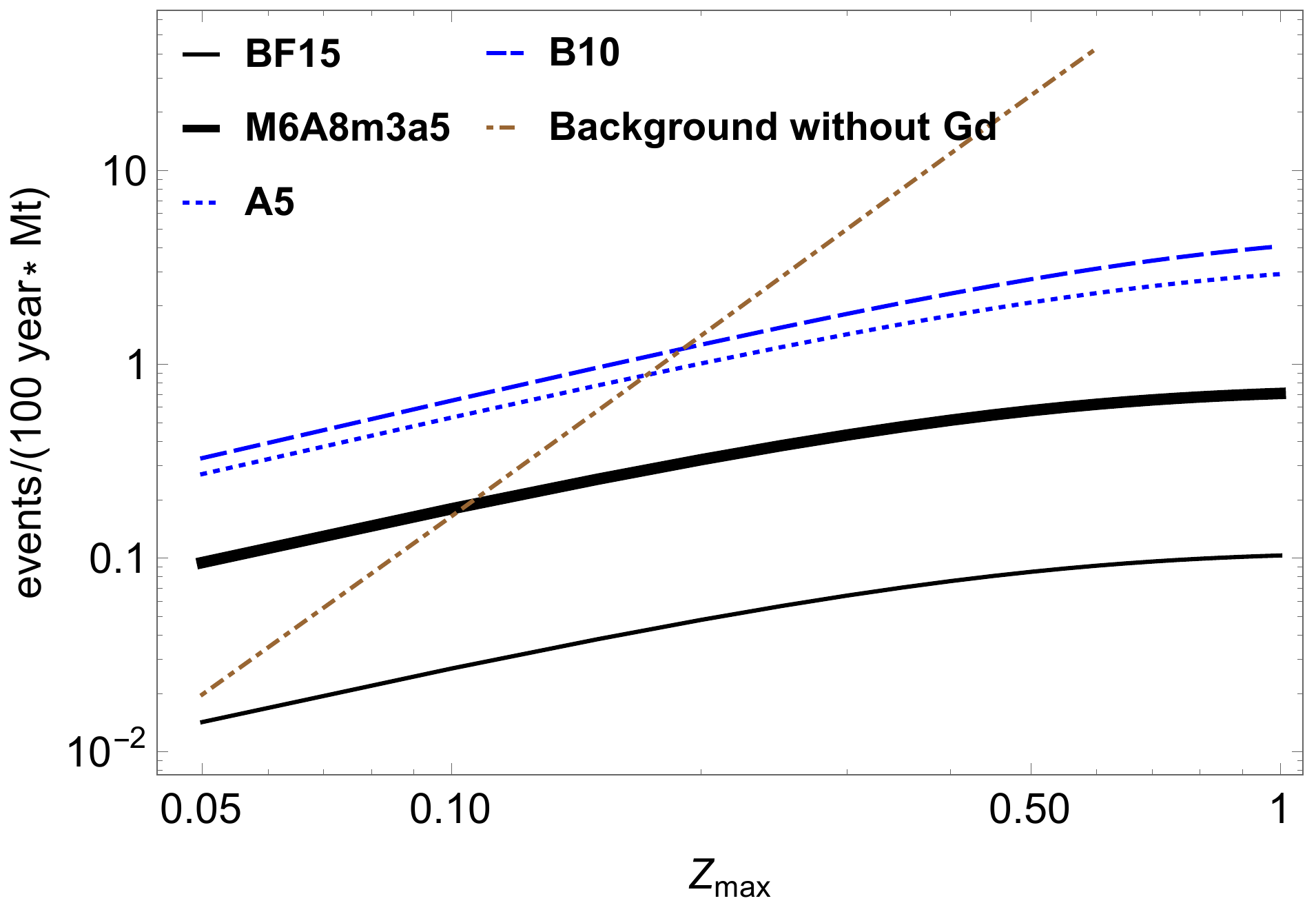} 
	\caption{The same as Fig. \ref{fig:ratesz}, for \nsbh\ mergers.   }
	\label{fig:rateszBH}
\end{figure*}

When searching for \ns\ in time coincidence with a \gw\ detection, a relevant question is the distance (or, equivalently, redshift) range of sensitivity of the \n\ detector.   Fig. \ref{fig:zbins} shows an example of distribution in redshift bins (of width $\Delta z=0.05$) of the number of mergers expected to be observed at Earth (assuming 100\% efficiency in \gw\ detection), and of the \n\ signal and background event rates.  The most optimistic models of \n\ emission (\nsns\ merger with long lived HMNS from model M, see Table  \ref{tab:models}) and of \mg\ rates were chosen here (see figure caption). Additionally, a detector with Gd was assumed here.

We see that the number of mergers grows rapidly with $z$ for $z<0.5$, reflecting the growth of the merger rate and of the cosmic volume with redshift.  For $z<0.05$ ($z<0.5$), $N\sim 10^4$ ($N\sim 10^6$) mergers per century are expected, in agreement with existing predictions for  \gw\  detectors of the next generation, see, e.g., \cite{Mills:2017urp}.  

The number of background events in a \n\ detector roughly follows the growth of the number of \mgs, with a faster increase with $z$ due to our accounting for time dilation when choosing the time window of observation for coincidence (see Sec. \ref{sub:method}).  The number of signal events can be as high as $N_s \sim 10$ for $z<0.05$. It decreases with $z$, because the increase of the merger rate with $z$ is overcompensated by effects of the redshift of energy (as the redshift increases, a larger fraction of the \n\ flux falls below the energy window), and of the  volume-dilution of the \n\ flux from a single source. 
The rate of signal events exceeds the background in the first 3-4 bins, corresponding to $z < z_{S/B}\simeq 0.2$, where $z_{S/B}$ is the redshift at which the number of background events becomes comparable to the number of signal events.
\begin{figure}[h]
	\centering
	\includegraphics[width=0.45\textwidth]{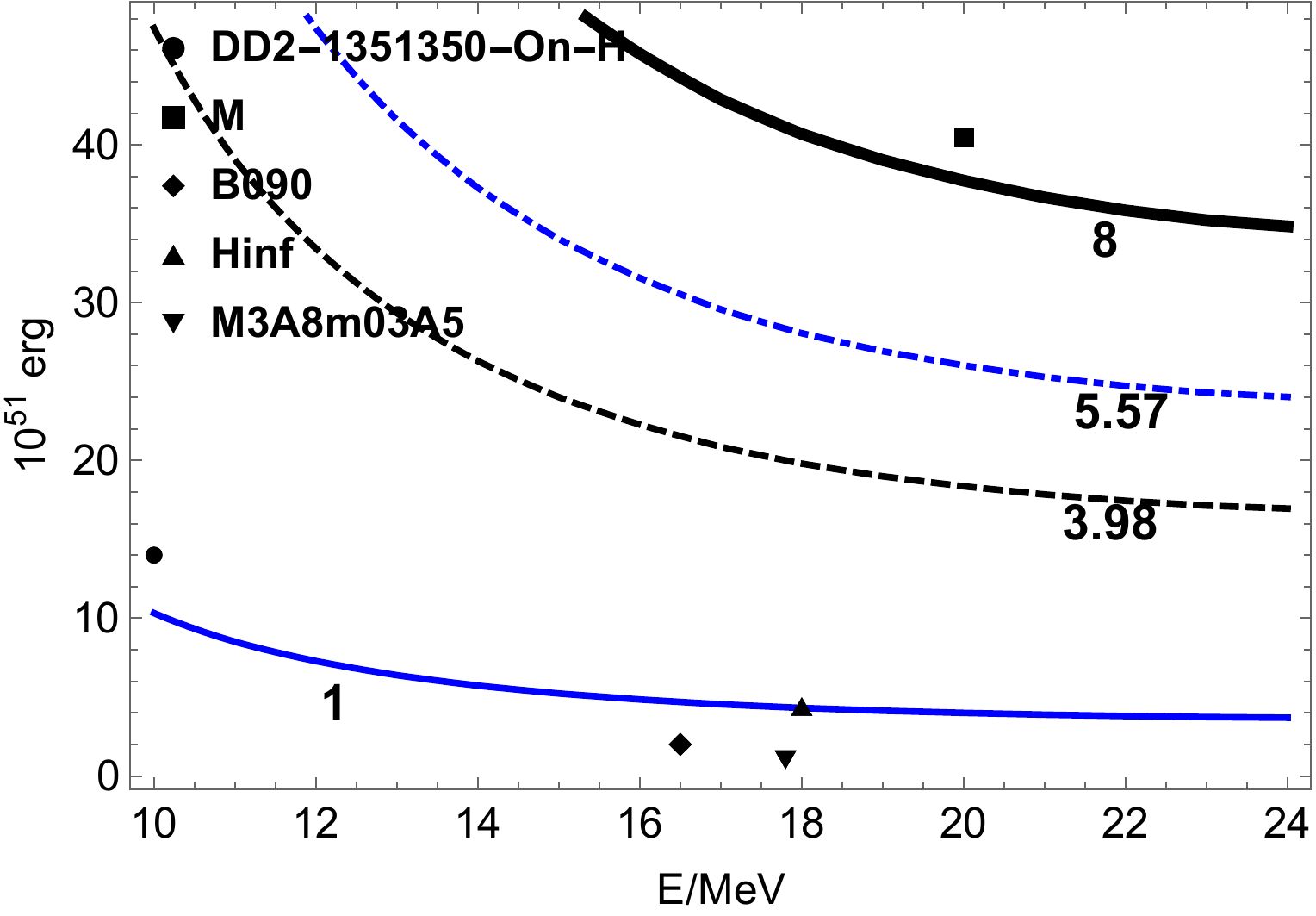} 
	\includegraphics[width=0.45\textwidth]{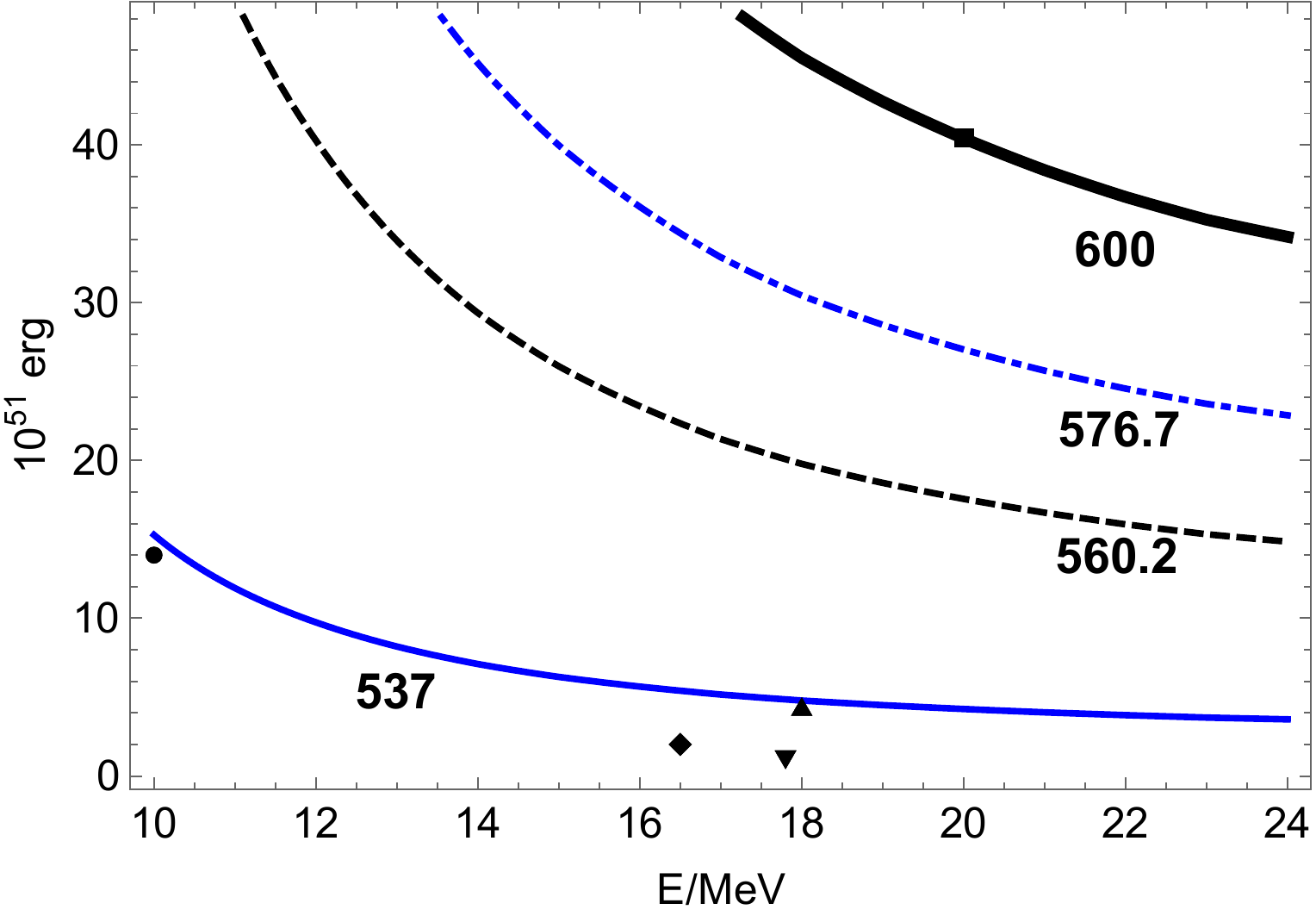} 
	\caption{
	Iso-contours of the total number of events (signal plus background, numbers on curves) from BNS mergers, in the space of the total energy and average energy of the emitted $\barnue$ of a single merger, for redshift intervals of $0<z<0.05$ (upper panel) and of $0<z<0.74$ (lower panel).  We assumed the optimistic merger rate, an exposure of 100 ${\rm  yr \cdot Mt}$, and a detector with Gd. Predictions from models of BNS mergers (see Table \ref{tab:models}) are shown for comparison.  The number of neutrino events from the background is 0.17 and 529 in the upper and lower panel, respectively.
%
%
The black dashed line represent the lower edge of a $90\%$ confidence interval centered at the solid thick black contour, while the blue dot-dashed line represents the upper edge of a $90\%$ confidence interval centered at the blue solid contour. 
		}
	\label{fig:Contour}
\end{figure}

Figs. \ref{fig:ratesz} and \ref{fig:rateszBH}  show the number of signal and background events for the interval $z< z_{max}$, as a function of $z_{max}$, for the different models considered in this work. Results are shown for different combinations of merger types, merger rates and detector configuration. 
For the case of water with Gd, the figures confirm the results shown in \figu{zbins} for the most optimistic scenario.  In addition, they show that for other models of \nsns\ \mgs, and for all the models of \nsbh\ \mg, the numbers of signal events are at the level of $N_s \sim 0.1-2$  for $z_{max}\sim 0.1$.   Backgrounds overwhelm the signal already at $z_{S/B} \sim 0.1-0.2$.  
For pure water, the figures reveal that the situation is noticeably worse due to the larger background rate.  $z_{S/B} < 0.1$ is expected for the more conservative flux models, and in certain cases the background exceeds the signal at all $z$.

\begin{figure}[h]
	\centering
	\includegraphics[width=0.45\textwidth]{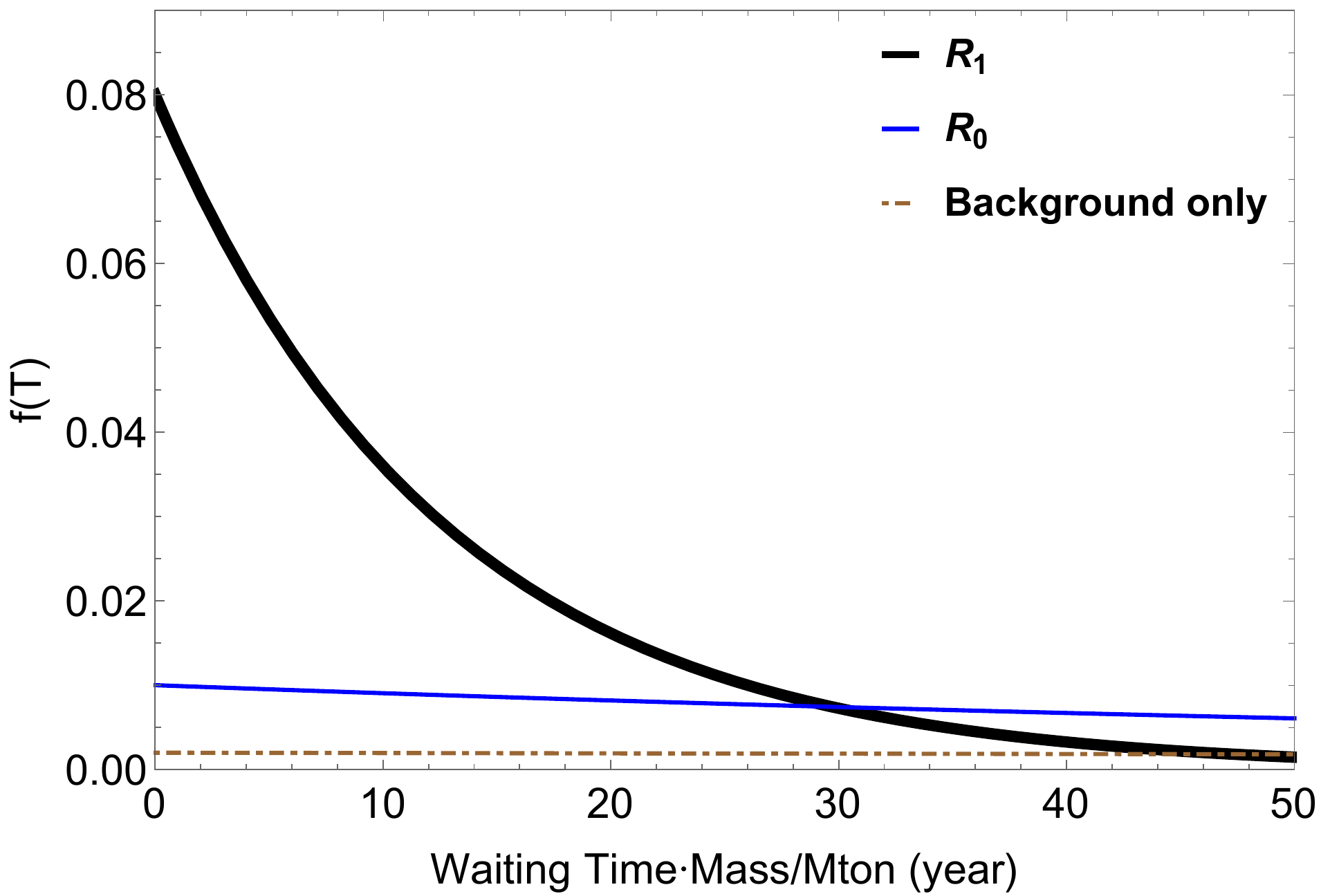} 
	\vskip 0.2truecm
	\includegraphics[width=0.45\textwidth]{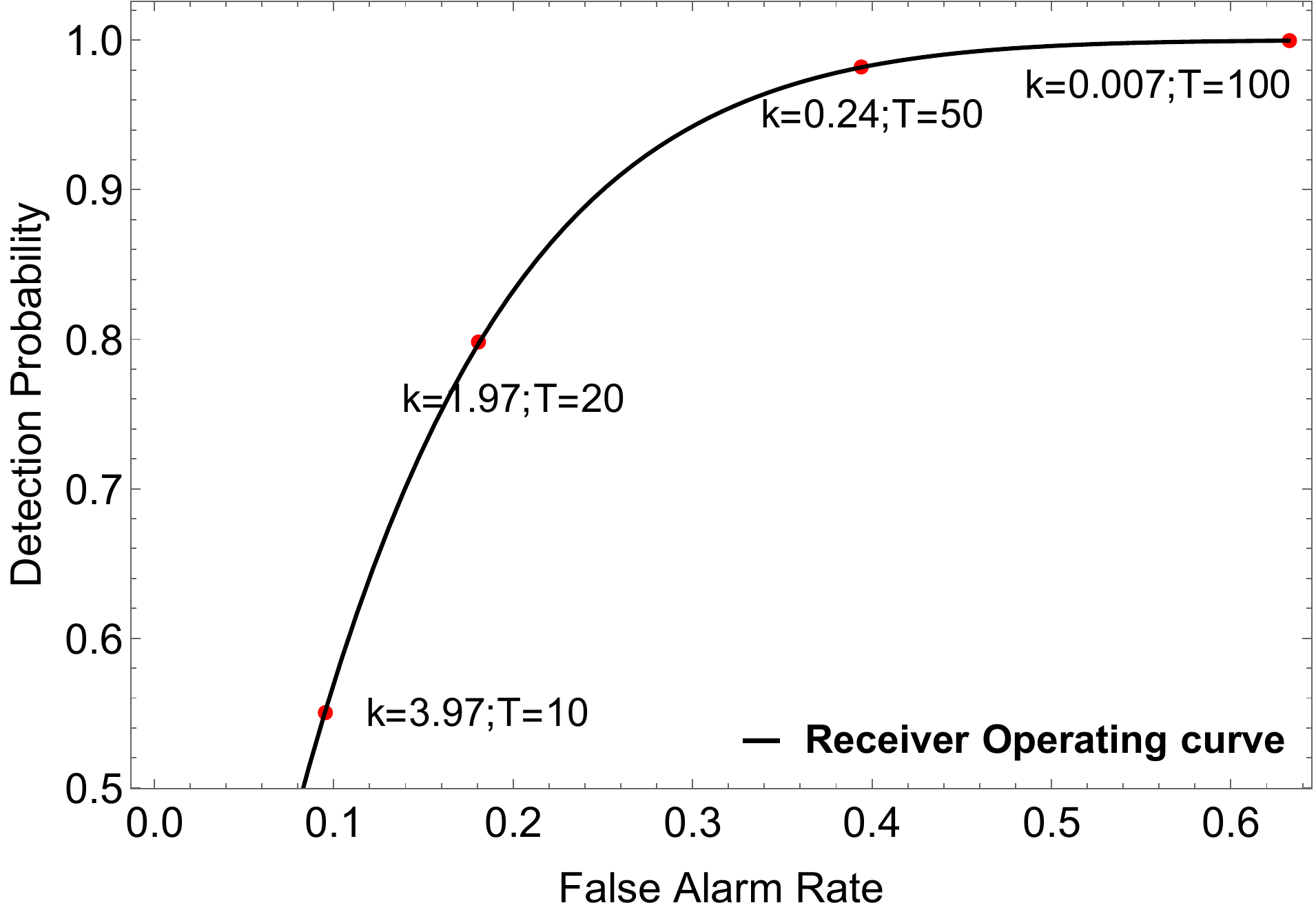} 
	\caption{ 
	{\it Upper panel:}
		the probability density of the waiting time to the first arrival, $f(T)$, 
		as a function of exposure, for two generic models of BNS mergers taken from Fig. \ref{fig:Contour} (solid thick black and solid blue curves in the upper panel of Fig. \ref{fig:Contour}). One model predicts a total event rate of $R_1=8 \cdot 10^{-2}~{\rm yr^{-1}}$; the other predicts $R_0= 10^{-2}~{\rm yr^{-1}}$ (see legend in figure). We assume  \nsns\ mergers with the optimistic merger rate \cite{Eldridge:2018nop}, and a 1 Mt detector with Gd. 
		The background-only case is shown for reference as well. %
%
		%
		{\it Lower panel:}
		 the probability $P_D$ of favoring the more optimistic model (the one with rate $R_1$), as a function of false alarm rate $P_F$.
		 For certain points on the curve (red dots) the corresponding values of the waiting time threshold $T_k$, in unit of years (or, equivalently, the likelihood ratio threshold, $k$) are shown. See text for more details.
		}
	\label{fig:waiting}
\end{figure}

 Let us briefly discuss the statistical significance of a detection. As a criterion for statistical significance we establish that the total predicted number of events, $N_{tot}=N_S + N_B$, should be higher than a threshold $N_T$ beyond which the hypothesis of background-only is unlikely to be realized:  $\sum_{n=N_T}^{\infty}P(n|N_B)< 0.0015$.
Here $P(n|N_B)$ is a Poisson distribution of $n$, with the mean being $N_B$. The criterion we apply here naturally becomes the 3-sigma-limit when the statistics is high.

The statistical significance of a detection is dependent on both the effective observation range, $z_{max}$, and the operation time of a network of neutrino/GW detectors. Because the signal-to-background ratio decreases with $z_{max}$, for a given flux model and exposure, one can find the maximum $z_{max}$ within which the signal could be distinguished from noise with high statistical significance. 
For example, for the scenario in the left upper (left bottom) panel of Fig. \ref{fig:ratesz}, where \nsns\ \mgs\ predicted by model M with optimistic (moderate) merger rate, a detector with Gd, and a $100$ Mt year exposure are used, 
we find that this extreme value is $z_{max}=0.74$ ($z_{max}=0.29$), for which the number of signal events is $N_S=71$ ($N_S=11$). The corresponding number of background events is $N_B=529$ ($N_B=10$). 
Including mergers with higher redshift in the analysis will worsen the significance of the signal, due to the overwhelming background.

Let us briefly discuss the physics potential of a search for \ns\ from mergers. Given a measured event rate, it will be possible to place constraints on the main parameters governing the merger \n\ flux, namely the total energy $\mathcal{E}_{\bar{\nu}_e}$ and the average energy of the $\barnue$ spectrum,$\langle {E_{\barnue} }\rangle$. As an illustration of possible constraints on these parameters, in Fig. \ref{fig:Contour} we show two sets of iso-contours of the total  number of  events (signal plus background) in the space of $\mathcal{ E}_{\barnue}$ and $\langle {E_{\barnue} }\rangle$, for the two redshift intervals of reference, $z\leq 0.05$ and $z\leq 0.74$.The optimistic BNS merger rate was adopted here. For comparison, the predictions from each of the models in Table \ref{tab:models} for BNS mergers are shown. We assume that uncertainties on the merger rates and on the background rates are negligible, so errors on the number of events are only statistical. 
In each of the panels of the figure, the contours are chosen so that they form two bands of $90\%$ confidence level statistical uncertainty  centered around two hypothetical measured event rates, one pessimistic and one optimistic ($N_{tot}=1,8$ for $z\leq 0.05$  and $N_{tot}=537, 600$ for $z\leq 0.74$, see figure caption for more information).  These bands represent the region of the parameter space that can be considered consistent with the measurement.   
We see that a measurement of a relatively large event rate would disfavor a wide portion of the parameter space, and result in a lower bound on ${\mathcal E}_{\barnue}$, under minimal assumptions on the average energy (e.g., if one assumes ${\mathcal E}_{\barnue } < 24$ MeV as a theoretical prior). In the case of a low measured number of events, most of the parameter space would be allowed (including ${\mathcal E}_{\barnue}=0$, the background-only case), and only the most extreme corner of the parameter space would be disfavored. In this latter case, most or even all the models would be consistent with the data.

In addition to bounds from the event rate, it will be possible to further constrain the parameter space using the energy distribution of the events. Roughly, these will correspond to vertical lines in Fig. \ref{fig:Contour}, which we do not show for simplicity. More realistically, a  fit of the data energy spectrum and overall normalization will give characteristic boomerang-shaped confidence level contours similar to those obtained for \ns\ from the supernova SN1987A (see, e.g. \cite{Lunardini:2000sw} and references therein). A detailed illustration of the potential of data fits is beyond the scope of the present work. 

\subsubsection{Probability of first detection}
\label{subsub:probdet}

As was shown in the previous section (see Fig. \ref{fig:Contour}), 
even the detection of a single candidate event, or even a null result (no detection) can be significant, and provide useful constraints. This can be especially relevant in the short term, when the detector exposure is still low.  As a further illustration, 
we follow Kyutoku and Kashiyama \cite{Kyutoku:2017wnb}, and discuss how the waiting time -- i.e., the time that elapses from the beginning of the experimental search until the first candidate event is observed -- is a useful statistical observable to distinguish between different merger models.  

We consider the total number $N_{tot}$ of events that are observed in time-coincidence with GW detections, and suppose that it follows a Poisson process with event rate $R$. 
Then the length of time until the first arrival, denoted by $T$, is a continuous random variable. In a simple scenario where $z_{max}$ is fixed (rather than adjusted depending on the model, to optimize the statistical significance), the rate is a constant, $R=N_{tot}/T$, which is determined by the merger flux model. The background rate can be assumed as known here.  

The cumulative distribution function of $T$, which is the probability that the first arrival is observed within the time interval $[0,T]$, is:
\beq
F(T)=1-\frac{(RT)^0\exp^{-RT}}{0!}=1-e^{-RT}~,
\eeq
and the probability density that the first arrival is observed within the time interval $[T,T+dT]$ is\footnote{Note that $f(T)dT$ is the probability that no events are recorded in the interval $[0,T]$, and one event is recorded in the interval $[T,T+dT]$.}:
\beq
f(T)=\frac{dF(T)}{dT}=R e^{-RT}.
\eeq
With a single measurement of $T$ (waiting time of the first event arrival) with probability density function given by $f(T)$, 
one can use a likelihood ratio test to distinguish between two hypotheses, $H_1$, and $H_0$, according to the Neyman-Pearson lemma \footnote{When we want to distinguish an alternative theory $H_1$ from a default (``null") theory $H_0$, a type I error occurs if we reject $H_0$ when it is true. The power of a hypothesis test is the probability of making a correct decision if $H_1$ is true.
	The Neyman-Pearson lemma \cite{kayfundamentals} states that  a likelihood ratio test is the test that has maximum power
	for a given  type I error. }.

To fix the ideas, let us consider 
an example drawn from Fig. \ref{fig:Contour} (upper panel): a search for \ns\ from BNS mergers, for the optimistic merger rate, a 1 Mt detector (with Gd) and $z_{max}=0.05$. In this example, hypothesis $H_1$ predicts a rate $R_1=8 \cdot 10^{-2}~{\rm yr^{-1}}$ (black solid curve in the upper panel of Fig. \ref{fig:Contour}), while $H_0$ predicts $R_0=10^{-2}~{\rm yr^{-1}}$ (blue solid curve in the same panel). 

%
Let us introduce the likelihood ratio: 
\beq
\Lambda(T)=\frac{f(T;R_1)}{f(T;R_{0})}=\frac{R_1}{R_{0}}e^{-(R_1 - R_{0})T}~. 
\label{eq:likratio}
\eeq
When a measurement gives $\Lambda \gg 1$ ($\Lambda \ll 1$), hypothesis $H_1$ ($H_0$) is strongly favored over the other one. Focusing on the case $\Lambda >1$ for definiteness, one can choose a threshold value, $k>1$, such that if $\Lambda >k$ hypothesis $H_1$ is accepted as true, and $H_0$ is discarded (as unlikely to be true). 

Let us also define the probabilities:
\beq
&P_D=\int_{T:\Lambda(T)>k}f(T;R_1)dT,\\
&P_F=\int_{T:\Lambda(T)>k}f(T;R_{0})dT,
\label{eq:pdpf}
\eeq
where the integration is performed in the interval $[0,T_{k}]$, where $T_k$ is the value of $T$ such that $\Lambda = k$ is satisfied. 

Here $P_D$ (``probability of detection", in common jargon) is the probability that hypothesis $H_1$ is correctly chosen. In other words, it is the probability that the analysis will correctly identify the true hypothesis, and represents the power of the hypothesis test. 
$P_F$ is a ``false alarm" probability; it is the probability that $H_1$ is (wrongly) chosen while $H_0$ is true. It represents the probability that the analysis will give an incorrect answer, and is a type I error (see, e.g. \cite{kayfundamentals}).  
Both $P_D$ and $P_F$ increase with decreasing likelihood ratio threshold $k$, or, equivalently, with increasing waiting time threshold $T_k$. A detector with good performance to test one model against another is expected to have high $P_D$ while low $P_F$. This behavior can be seen in the Receiver Operating Characteristic (ROC) curve, which is defined as the path in the space of $P_F$ and $P_D$ which is obtained by varying $k$.

For our example, the ROC curve is shown in the bottom panel of Fig. \ref{fig:waiting}.   From this figure we learn that, if a relatively short waiting time threshold is set, e.g., $T_k=10$ yr, the power of the test is moderate, $P_D\sim 0.55$, although the type I error associated to it is rather low, $P_F \simeq 0.1$. Qualitatively, this can be understood by thinking that, if a short arrival time is observed, $T<T_k=10$ yr, then $H_1$ will be favored with a good level of confidence, but the probability of this situation to be realized is relatively low if $H_0$ is true.  

If a  longer waiting time threshold is established, e.g. $T_k=20$ yr, the power of the test increases to $P_D\simeq 0.82$, however the  probability to falsely choose $H_1$ when instead $H_0$ is true also increase, $P_F \simeq 0.24$. Intuitively, this means that, by setting a weaker criterion on the likelihood ($\Lambda \geq 1.52$), we obtain a test which is more likely to favor $H_1$ (because the condition on the likelihood is rather likely to be met), however this comes at the price of having a larger probability to come to the incorrect conclusion. 

If a very long waiting time is observed, $T\sim 100$ yr, then $H_0$ will be strongly favored over model $H_1$, because $\Lambda(T)\ll 1$, and for even larger $T$ both models might be poor interpretations of the data, and a third model (e.g., background only) might have to be considered as an alternative.  Indeed, we must emphasize that the likelihood ratio method discussed here is in the spirit of discriminating between two models only, and therefore might be suitable to use if the theoretical panorama evolves towards a somewhat bipolar situation (two fairly established classes of models only).

\section{Summary and discussion}
\label{sec:disc}

We found that the detection of thermal \ns\ from \mgs\ is possible, over the next several decades,  if the synergy between Mt-scale \n\ detectors and next generation \gw\ detectors, capable of observing \mgs\ up to redshift $z\sim 2$, is fully exploited.  

A very important element is the time coincidence between \gw\ and \n\ detection, which will have two important benefits. The first is a strong reduction of the background at the \n\ detector, without which the identification of the \n\ flux from \mgs\ would be almost impossible. The second is that each \n\ candidate  event can be analyzed as a single, stand-alone signal, with measured energy and arrival time, and with an identified (potential) parent \mg, of known \mg\ time, redshift, type (\nsns\ or \nsbh) and morphology. 

About the detectability of \ns\ from \mgs, 
we find that, among the \n\ emission models we have considered, the strongest \n\ 
signal is expected for \nsns\ \mgs\ (see Table   \ref{tab:models}), 
if their rate is in the higher part of the interval currently allowed by the \lgvg\ data, and if the compact object emits more than $\sim 10^{52}$ ergs in $\barnue$s, due to its being a long-lived or hypermassive neutron star. 
In this scenario, up to ${\mathcal O}(10^2)$ \n\ events  are predicted for an exposure of $100~{\rm Mt\cdot yr}$, for an energy window $E_{e}=10-50$ MeV and a redshfit window $z \simeq 0 - 0.5$. 
Several tens of percent of the candidate events (those with $z\lesssim 0.1$ or so) will be attributed to parent mergers of \emph{known redshift} with high statistical significance (see Fig. \ref{fig:zbins}). And part of them might even be \emph{individually}  statistically significant over the background.

For  \mgs\ with more conservative parameters, i.e. lower \mg\ rates, and lower total \n\ luminosity --  $E_{\barnue }\sim few \cdot 10^{51}$ ergs -- 
the event rates can be ${\mathcal O}(1)$ event or less for the same exposure ($100~{\rm Mt\cdot yr}$), and the minimum exposure needed for a first detection would be several tens of ${\rm Mt\cdot yr}$. Because the parameters and models considered here are only a subset of all the possibilities, our results should be interpreted as a broad (and tentative) interval of possible \n\ detection rates: $\dot N \sim (10^{-3} - 1)~{\rm Mt^{-1}yr^{-1}}$. 

On the front of what can be learned about \mgs\ from a \n\ detection, data from a \n\ detector --  after the background reduction allowed by the time-coincidence method -- will probably be analyzed statistically to extract basic, low precision, information on the merger \n\ flux parameters, mainly the energy spectrum and the total energy emitted. These will then be used to constrain  numerical models of \n\ emission, thus indirectly testing the features (e.g., mass of the compact object and/or of the accretion disk) of the underlying merger models. 
The methods used in such analyses will draw from the experience of other high-background searches, for example those for the diffuse supernova neutrino background, where the energy distribution and total count of the data are fit with a mixed model (signal plus background) with both signal and background parameters to be found by fitting (see, e.g. \cite{Malek:2002ns}). Advanced analyses where the time window $\Delta t_i$ is also optimized by fitting may further constrain the underlying model, and also enhance the signal significance by further reducing the background contamination. 

In the fortunate case (see above) where a few single \n\ detections are individually significant, there might be potential to search for correlations with the physical features of the mergers provided by the \gw\ observation, using methods similar to those used at IceCube in analyses of high energy \n\ data in a multi-messenger context (see, e.g.  \cite{IceCube:2018dnn}).  A full exploration of the potential of such event-by-event analyses is left for future work.

We emphasize that even a single \n\ detected in coincidence with a \mg, or even a non-detection, can place important constraints on the neutrino luminosity in a \mg, and therefore on the type and physics of the newly born compact object, if the detector exposure exceeds $\sim$10 Mt yr for a configuration with water plus Gadolinium. In this context, the waiting time before the first detection is a useful statistical observable.  Considering 
HyperKamiokande (in its futuristic upgrade to water plus Gd, with the currently planned fiducial mass of $\sim 370$ kt)  a waiting time of $
\sim$30 years is needed to start constraining theoretical flux models. Such waiting time is realistic for a long-term, broad-scope project like HyperKamiokande.

In closing, we believe that the detection of thermal \ns\ in time-coincidence with gravitational waves from mergers is a new, realistic goal to be added to the agenda of Mt-scale \n\ detectors with $\sim$ MeV energy thresholds. It may be the first observation of $\sim 10$ MeV-scale \ns\ from individually resolved cosmological sources, thus drastically expanding the horizon of possibilities of low energy \n\ astronomy in the context of multi-messenger studies.

\subsection*{Acknowledgments}

The authors acknowledge funding from the National Science Foundation grant number PHY-1613708. They thank the authors of ref. \cite{Lippuner:2017bfm} for sharing their numerical results and for discussions. They are also grateful to Liliana Caballero, Laura Cadonati and Mohammadtaher Safarzadeh for useful comments. 



 \bibliography{b}

\end{document}